\documentclass[a4paper,13pt]{article}
\usepackage[english]{babel}
\pdfoutput=1 % if your are submitting a pdflatex (i.e. if you have
\usepackage{jheppub} % for details on the use of the package, please
\usepackage[T1]{fontenc} % if needed
\usepackage[utf8]{inputenc}

\usepackage{comment}

\usepackage{cases}

\newcommand{\bi}{\begin{itemize}}
    \newcommand{\ei}{\end{itemize}}
\newcommand{\bea}{\begin{eqnarray}}
    \newcommand{\eea}{\end{eqnarray}}
\newcommand{\bt}{\begin{tabular}}
    \newcommand{\et}{\end{tabular}}
\newcommand{\bc}{\begin{center}}
    \newcommand{\ec}{\end{center}}

\newcommand{\be}{\begin{equation}}
    \newcommand{\ee}{\end{equation}}
\newcommand{\ba}{\begin{array}}
    \newcommand{\ea}{\end{array}}

\def\bbox{{\,\lower0.9pt\vbox{\hrule \hbox{\vrule height 0.2 cm
                \hskip 0.2 cm \vrule height 0.2 cm}\hrule}\,}}
\newcommand{\dsl}{\pa \kern-0.5em /}

\makeatletter \@addtoreset{equation}{section} \makeatother

\def\slashchar#1{\setbox0=\hbox{$#1$}           % set a box for #1
    \dimen0=\wd0                                 % and get its size
    \setbox1=\hbox{/} \dimen1=\wd1               % get size of /
    \ifdim\dimen0>\dimen1                        % #1 is bigger
    \rlap{\hbox to \dimen0{\hfil/\hfil}}      % so center / in box
    #1                                        % and print #1
    \else                                        % / is bigger
    \rlap{\hbox to \dimen1{\hfil$#1$\hfil}}   % so center #1
    /                                         % and print /
    \fi}

\pdfoutput=1

\title{\boldmath   $\mathcal{N}=2$ conformal higher-spin actions in harmonic superspace}

\author[a]{Evgeny~Ivanov,}
\author[a,b]{Nikita~Zaigraev}

\affiliation[a]{Bogoliubov Laboratory of Theoretical Physics, JINR,\\
    Joliot-Curie st. 6, 141980 Dubna, Moscow region, Russia}
\affiliation[b]{Moscow Institute of Physics and Technology,\\ Institutskiy per. 9, 141700 Dolgoprudny, Moscow region, Russia}
\emailAdd{eivanov@theor.jinr.ru}
\emailAdd{nikita.zaigraev@phystech.edu}

\abstract{
    We construct linearized $4D, \mathcal{N}=2$ Weyl-type off-shell  actions in harmonic superspace for the superconformal higher-spin gauge multiplets
    introduced in \href{https://arxiv.org/abs/2404.19016}{arXiv:2404.19016}. This work is a followup
    of our \href{https://arxiv.org/abs/2511.16325}{arXiv:2511.16325}, where an analogous construction was carried out for conformal $\mathcal{N}=2$ supergravity.
    We check the superconformal invariance of the new actions and their linearized gauge invariance.
    As the characteristic features of our formulation, the structure of the underlying prepotentials and their gauge and superconformal transformations are entirely determined
    by the fundamental \textit{principle of Grassmann harmonic analyticity}, while the construction of the invariant actions and of the generalized $\mathcal{N}=2$ super-Weyl tensors relies
    upon the \textit{principle of Grassmann half-analyticity}. This provides not only a new geometric framework for the linearized
    $\mathcal{N}=2$ superconformal gauge higher-spin theories, but also suggests possible ways of promoting them  to the full nonlinear level.
    We  consider a generalization to an arbitrary $\mathcal{N}=2$ supergravity background
    and show that the linearized spin-$\mathbf{s}$ gauge invariance, without radical modifications of the gauge transformation law and/or harmonic zero-curvature equations,
    is consistent only with $\mathcal{N}=2$  Weyl-flat super-backgrounds.
 }

\makeatletter
\gdef\@fpheader{}
\makeatother

\begin{document}

\maketitle
\flushbottom

%\newpage

\section{Introduction}

Conformal higher-spin field theories form an important class of gauge models that naturally generalize the linearized Weyl gravity to fields
with arbitrary integer or half-integer spins \cite{Fradkin:1985am, Metsaev:2007rw, Vasiliev:2009ck}.
%These
%theories were further developed by Fradkin and Linetsky ,
The relevant algebras and cubic vertices, including the superconformal case, were constructed in \cite{Fradkin:1989md, Fradkin:1989yd, Fradkin:1990ps, Vasiliev:2001zy}.
The subsequent evolution of conformal higher-spin theory was largely
shaped by the seminal works \cite{Tseytlin:2002gz} and \cite{Segal:2002gd}. There, a unique non-linear bosonic gauge theory accommodating interacting,
symmetric, traceless tensor fields of
arbitrary rank in any even dimension $d \geq 4$ was presented. This framework is now widely known
as conformal higher-spin gravity \cite{Bekaert:2010ky, Bekaert:2022poo, Beccaria:2017nco, Basile:2022nou}.

\smallskip

Since the successful superfield construction of free $\mathcal{N}=1$ superconformal actions in ref. \cite{Kuzenko:2017ujh}, the superfield approach
to higher-spin gauge theories has undergone rapid and extensive development.
There were thoroughly investigated the conformal geometry of such supersymmetric gauge theories \cite{Kuzenko:2019ill}, formulated new locally superconformal
gauge models in Bach-flat backgrounds \cite{Kuzenko:2020jie},
proposed generalized superconformal higher-spin multiplets \cite{Kuzenko:2020opc}, explored the conformal interactions between higher-spin gauge
superfields and matter \cite{Kuzenko:2022hdv}, and calculated induced
superconformal higher-spin actions to the quadratic order \cite{Kuzenko:2022qeq}. In recent years, the considerable attention was directed towards
generalizing these advancements  to the case of extended supersymmetry \cite{Kuzenko:2021pqm, Hutchings:2023iza, Kuzenko:2024vms}.

\smallskip

The most natural and powerful framework to deal with unconstrained off-shell $\mathcal{N}=2$ supersymmetric theories
is the harmonic superspace (HSS) approach \cite{Galperin:1984av, 18}. By augmenting the standard superspace with additional harmonic variables,
HSS allows for a clear geometric formulation of supersymmetric gauge theories founded on the \textit{principle of harmonic Grassmann analyticity}\cite{GRAn}. In recent years,
this approach was successfully applied to describing massless $\mathcal{N}=2$ higher-spin gauge multiplets  \cite{Buchbinder:2021ite} and
their interactions \cite{Buchbinder:2022kzl, Buchbinder:2022vra}. Building on these developments, the geometric off-shell
formulation of $\mathcal{N}=2$ superconformal higher-spin gauge multiplets was recently given in \cite{Buchbinder:2024pjm}
(see also  \cite{Buchbinder:2024xll, Buchbinder:2025ceg}). The corresponding higher-spin analytic prepotentials
were introduced as analytic connections in the harmonic derivative $\mathcal{D}^{++}$, with values in the proper subalgebras of the enveloping algebra
of derivatives with respect to the analytic-basis HSS coordinates.  However,
the explicit construction of the dynamical gauge-invariant actions in HSS  for these multiplets (the higher-spin super Weyl actions) remained an open problem.

The primary goal of the present paper is to fill this gap. We set up the HSS $\mathcal{N}=2$
superconformal actions for the higher-spin multiplets introduced in \cite{Buchbinder:2024pjm}. To achieve this, we follow the non-conformal case \cite{Ivanov:2024gjo, Ivanov:2025mld}
and introduce supersymmetry-invariant (though non-analytic) $G^{++}$-superfields to define the \textit{half-analytic} gauge-covariant
$\mathcal{H}^{++}_{\alpha(2s-2)}$ superfields. From these objects we compose the gauge-invariant $\mathcal{N}=2$ super Weyl tensors
and the relevant Weyl actions for an arbitrary integer spin $\mathbf{s}$.
A generalization to the half-integer spins, which could involve a different multiplet structure with spinor analytic prepotentials \cite{Ivanov:2024bsb},
is deferred to a separate consideration. The crucial point of our analysis is the explicit verification
of $\mathcal{N}=2$ superconformal invariance of the  actions constructed, which required a non-trivial adaption
of the methods employed in the linearized $\mathcal{N}=2$ conformal supergravity \cite{Ivanov:2025gvs}.

\smallskip

Several important aspects are to be highlighted before passing to the detailed exposition. Actions for
$\mathcal{N}=2$ superconformal higher-spin theories in superspace were earlier constructed
in the framework of $\mathcal{N} = 2$ ``conformal superspace'' approach \cite{Kuzenko:2021pqm}.
One of the advantages of the latter is the manifest superconformal symmetry. Nevertheless,
to be aware of the analogous actions in harmonic superspace is of  significant interest for a number of reasons.

\smallskip

\textbf{1.} As discussed in \cite{Buchbinder:2024pjm}, the gauge prepotentials introduced in \cite{Kuzenko:2021pqm}
can be  recovered by fixing a special harmonic-independent gauge. Thus, the conformal superspace formulation
can naturally be \textit{embedded into the harmonic framework}. It should be emphasized that both rigid and local ${\cal N}=2$
superconformal symmetries are already contained in the general analyticity-preserving diffeomorphisms of ${\cal N}=2$ HSS, before imposing any gauge.

\textbf{2.} The harmonic formulation provides an exceptionally powerful
and convenient toolkit for quantum computations, particularly for \textit{evaluating induced effective actions}.
The well-established covariant supergraph techniques in HSS allows one to derive higher-spin superconformal
invariants directly by integration over hypermultiplet matter in the relevant functional integral. Some first steps towards this goal
were undertaken in~\cite{Buchbinder:2025ceg}.

\textbf{3.} The fundamental advantage of the HSS approach is its \textit{intrinsically geometric nature}.
Within this framework, the higher-spin gauge prepotentials naturally appear  as analytic connections, endowing
the free gauge transformations with a clear geometric interpretation. This geometric perspective is
\textit{invaluable for extending the theory beyond the free level}. Specifically, it provides a systematic pathway to set up interacting cubic vertices through
some natural non-Abelian deformations of the free analyticity-preserving gauge group, a task that
in the component or the standard superfield formalisms often requires a cumbersome guesswork.

\smallskip

It is useful to apply first to the non-conformal case in order to get insight into the difficulties that arise when passing to conformal models.
It is known how to construct super Weyl tensors for non-conformal $\mathcal{N}=2$ higher spins in HSS.
They are obtained as a straightforward generalization of the spin-$\mathbf{2}$ Weyl tensor \cite{Ivanov:2024gjo, Ivanov:2025mld, Ivanov:2025gvs}.
These tensors were actively used for finding out higher-spin $\mathcal{N}=2$ supercurrents and the corresponding
cubic vertices \cite{Zaigraev:2024ryg, Zaigraev:2025xqo, Zaigraev:2026jvu, Zaigraev:2026xzr, Zaigraev:2026vxf}. Nevertheless, attempts of their straightforward generalization
to the conformal case encounter several difficulties.
The primary challenge stems from the fact that the superconformal multiplet is described by a larger set of analytic prepotentials compared
to its non-conformal counterpart. Though these additional prepotentials can be gauged away in the Wess-Zumino gauge and so carry
no physical degrees of freedom at the end, their exact role in ensuring the correct structure of off-shell superfield actions and their
necessity  for $\mathcal{N}=2$ superconformal invariance remained highly non-obvious.
We will systematically address these challenges in this paper.

\smallskip

The paper is organized as follows. In Section \ref{sec: multiplets}, we briefly recall the procedure of setting up
$\mathcal{N}=2$ superconformal higher-spin analytic prepotentials and introduce the corresponding supersymmetry-invariant $G^{++}$-superfields.
Section \ref{sec: spin-3 theory} contains the details of $\mathcal{N}=2$ spin-$\mathbf{3}$ theory
and the spin-$\mathbf{3}$ super Weyl tensor. In Section \ref{sec: spin s theory}, we generalize this consideration
to an arbitrary integer spin $\mathbf{s}$. Section \ref{sec: superconformal transformations}
is devoted to the derivation of rigid superconformal transformations of the spin-$\mathbf{s}$ $G^{++}$-superfields
and to the explicit proof of invariance of the new higher-spin actions presented.
Finally,  Section \ref{sec: sugra back} contains a generalization of these actions to $\mathcal{N}=2$ supergravity background.
Section  \ref{eq: sec concl} collects the final comments and some problems for the future work.  In Appendix \ref{App: notations} we outline the notations
and conventions used over the paper.
In appendix \ref{App: half-analytic}  we show
that the superconformal gauge parameter can be chosen half-analytic.

\section{$\mathcal{N}=2$ superconformal higher-spin multiplets}\label{sec: multiplets}

In harmonic superspace, $\mathcal{N}=2$ superconformal higher-spin multiplets admit a concise geometric description \cite{Buchbinder:2024pjm}.
The corresponding analytic prepotentials arise as analytic connections in the harmonic derivative,
\begin{equation}
    \mathcal{D}^{++}
    \qquad
    \Rightarrow
    \qquad
    \mathcal{D}^{++} + \kappa_s \hat{\mathcal{H}}^{++}_{s} (J)^{P(s)},
\end{equation}
such that they take values in the proper finite-dimensional subalgebras of the enveloping algebra of derivatives in the coordinates of the analytic harmonic superspace.
The precise construction of such connections relies on few simple principles.
\smallskip

\noindent \textbf{1.} We consider a general analytic hypermultiplet vertex $(\mathbf{s}, \mathbf{\frac{1}{2}}, \mathbf{\frac{1}{2}})$,
motivated by the harmonic construction of the higher-spin hypermultiplet non-conformal vertices  \cite{Buchbinder:2022kzl}
\begin{equation}\label{eq: anz 1 st}
    S \sim \int d \zeta^{(-4)} \,
    q^{+a} h^{++M_{s-1}\dots M_1} \partial_{M_1} \dots \partial_{M_{s-1}} (J)^{P(s)} q^+_a,
    \qquad
     q^{+ a} = (\tilde{q}^+, q^+)\,,
     \quad
     q^+_a = \epsilon_{ab}     q^{+ b} .
\end{equation}
Here, $d\zeta^{(-4)}$
is the standard analytic superspace measure and $q^{+a}$
stands for  the off-shell hypermultiplet.
We also introduce the Pauli-G\"ursey breaking generator for odd values of $s$
\begin{equation}
    J q^+ := - i q^+,
    \qquad
J\tilde{q}^+ := i \tilde{q}^+,
\qquad
P(s) := \frac{1-(-1)^s}{2}.
\end{equation}
For even spins $P(s)=0$ and the generator $J$ drops out. This explicitly distinguishes
the interaction structures for even and odd spins, reflecting the fact that odd-spin gauge transformations essentially involve imaginary unit
(similarly to electromagnetic $U(1)$ symmetries), while even-spin transformations do not (analogously to gravitational translations).

Throughout the paper, we employ the standard definitions of the multi-indices and superspace derivatives:
\begin{equation}
    M \in \left\{ \alpha\dot{\alpha}, \alpha+, \dot{\alpha}+, ++ \right\},
    \qquad
    \text{and}
    \qquad
    \partial_M := \{ \partial_{\alpha\dot{\alpha}}, \partial^-_{\alpha}, \partial^-_{\dot{\alpha}},  \partial^{--} \}.
\end{equation}

\noindent \textbf{2.} Next, we require the invariance of the vertex \eqref{eq: anz 1 st} under rigid $\mathcal{N}=2$ superconformal transformations,
which act on the hypermultiplet in the well-known way.
By adjusting the form of the transformations for the introduced analytic prepotentials $h^{++\dots}$
and the necessary corrections to the operator $h^{++M_{s-1}\dots M_1} \partial_{M_1} \dots \partial_{M_{s-1}}$
we achieve $\mathcal{N}=2$ superconformal invariance of the resulting
model\footnote{We give a detailed discussion of the superconformal transformations in Sect. \ref{sec: superconformal transformations}.}.
 This allow us to fix the form of the higher-spin connection as
    \begin{equation}\label{eq: sc spin s operator}
    \hat{\mathcal{H}}^{++}_{s} := h^{++M_1\dots M_{s-1}} \partial_{M_{s-1}} \dots \partial_{M_1}
    +
    h^{++M_1\dots M_{s-3}}  \partial_{M_{s-3}} \dots \partial_{M_1}
    +
    \dots.
 \end{equation}
 Crucially, in addition to the leading term with
 $s-1$ derivatives, certain sub-leading terms containing
$ s-3$,
 $s-5$ derivatives,  etc., naturally emerge.
 At the same time, the contributions from the superconformal variation with
 $s-2$,
 $s-4$, ... derivatives can be absorbed into the terms already present in the operator.
 The technical details of this procedure were given in \cite{Buchbinder:2024pjm}.

Since the cubic vertex is defined as an integral over analytic superspace, the  higher-spin operator constructed in this way automatically
satisfies the analyticity condition
 \begin{equation}\label{eq: an cond}
    [\mathcal{D}^+_{\hat{\rho}},    \hat{\mathcal{H}}^{++}_{s}] = 0.
 \end{equation}
 Here, $\hat{\rho} = (\rho, \dot{\rho})$ denotes a collective spinor index ensuring the uniform description of objects with both chiralities.

The resulting  hypermultiplet model is given by:
\begin{equation}\label{eq: hyper action + sc}
    S_{sc} = - \int d\zeta^{(-4)} q^{+a} \left( \mathcal{D}^{++} + \kappa_s \hat{\mathcal{H}}^{++}_{s} (J)^{P(s)} \right) q^+_a.
\end{equation}
To investigate the gauge freedom of this action, we define the operator
\begin{equation}\label{eq: U oper}
    \hat{\mathcal{U}}_k := \left\{ \lambda^{M_1\dots M_{k-1}}, \partial_{M_{k-1}} \dots \partial_{M_1} \right\}_{\text{AGB}}\,,
\end{equation}
where the ``anti-graded bracket'' AGB is defined as
\begin{equation}
    \{B_1,B_2 \}_{\text{AGB}} := \{B_1,B_2 \},
    \qquad
        \{B,F \}_{\text{AGB}} := \{B,F \},
        \qquad
            \{F_1,F_2 \}_{\text{AGB}} := [F_1,F_2].
\end{equation}
Here $B$ stands for bosonic objects, and $F$ for fermionic ones.
In the simplest $k=2$ case, the AG bracket takes the
form\footnote{ The number $\epsilon_M \in \{0, 1\}$ denotes the Grassmann parity of the index $M$, {\it viz.} $0$
for bosonic coordinates and $1$ for fermionic ones.}
\begin{equation}
    \hat{\mathcal{U}}_2
    =
     \lambda^M \partial_M + (-1)^{\epsilon_M} \partial_M   \lambda^M
    =
    2 \lambda^M \partial_M + (-1)^{\epsilon_M} \left( \partial_M   \lambda^M \right),
\end{equation}
which is a sum of super translations and the weight transformation.
It is precisely through such an operator local $\mathcal{N}=2$ superconformal transformations act on the hypermultiplet.

We emphasize that the AG bracket is required here in order to properly account for the mixed Grassmann parities
of both the superspace derivatives and the corresponding gauge parameters. One of the main merits of the operator
introduced in this way is that it satisfies the ``intertwining'' relation
\begin{equation}
    \int d\zeta^{(-4)} P^{(q)} \, \hat{\mathcal{U}}_k  \, (J)^{P(s)}
     S^{(4-q)}
     =
     -  \int d\zeta^{(-4)} \,
     \hat{\mathcal{U}}_k \, (J)^{P(s)}
     P^{(q)}
     S^{(4-q)}
\end{equation}
for arbitrary analytic superfields $P^{(q)} $ and $ S^{(4-q)}$. Then, taking the gauge transformations of the hypermultiplet in the form
\begin{equation}
    \delta_\lambda^{(k)} q^+_a = - \kappa_s\, \hat{\mathcal{U}}_k   (J)^{P(s)} q^+_a,
    \qquad
    k = s, s-2, \dots\,,
\end{equation}
 we find that the spin-$\mathbf{s}$ analytic operator \eqref{eq: sc spin s operator} is defined modulo gauge transformations, which can be written in the form
    \begin{equation}\label{eq: sc spin s transformations}
    \delta^{(s,k)}_\lambda  \hat{\mathcal{H}}^{++}_{s} =
    \frac{1}{2} [\mathcal{D}^{++}, \hat{\mathcal{U}}_k ],
    \qquad
    k=s,s-2, \dots\,.
\end{equation}
Such transformations are parametrized by an additional integer $k$, which accounts for the gauge freedom of terms of different
orders in the superdifferential operator \eqref{eq: sc spin s operator}.
The value $k=s$ corresponds to the leading higher-spin gauge contribution, whereas $k=s-2, s-4, \dots$ label sub-leading gauge parameters
that secure the gauge freedom of the lower-order derivative terms in the operator \eqref{eq: sc spin s operator}.
The resulting variation \eqref{eq: sc spin s transformations} should be understood as a differential  operator acting on the hypermultiplet
and taking account of the analyticity of the latter. With this in mind and integrating by parts, we can cast this gauge variation in the more convenient form:
\begin{equation}\label{eq: transf conv}
    \delta^{(s,k)}_\lambda \hat{\mathcal{H}}^{++}_s
    =
    [\mathcal{D}^{++}, \hat{\Lambda}_s^{(k)}],
\end{equation}
where the analytic differential operator $\hat{\Lambda}_s^{(k)}$ is given by
\begin{equation}\label{eq: Lambda k}
    \hat{\Lambda}_s^{(k)} = \lambda^{M_1\dots M_{k-1}} \partial_{M_{k-1}} \dots \partial_{M_1}.
\end{equation}

\subsection*{Supersymmetry-invariant superfields}

Under the action of rigid $\mathcal{N}=2$ supersymmetry, the operator \eqref{eq: sc spin s operator}  is invariant;
however, because the derivatives $\partial_M$ transform nontrivially, the analytic prepotentials $h^{++}$
also have nontrivial transformation laws under both ${\cal N}=2$ Poincar\'e
and conformal supersymmetries.

In what follows, it will be convenient to pass to the covariant basis:
\begin{equation}
    \mathcal{D}_M : = \{\partial_{\alpha\dot{\alpha}}, \mathcal{D}^-_{\hat{\alpha}}, \mathcal{D}^{--} \},
\end{equation}
where the explicit expressions for the covariant derivatives in the analytic basis are given in Appendix~\ref{App: notations}.
In the covariant basis, the operator \eqref{eq: sc spin s operator} is rewritten as
        \begin{equation}\label{eq: sc spin s operator cov}
        \hat{\mathcal{H}}^{++}_{s} = G^{++M_1\dots M_{s-1}} \mathcal{D}_{M_{s-1}} \dots \mathcal{D}_{M_1}
        +
        G^{++M_1\dots M_{s-3}}  \mathcal{D}_{M_{s-3}} \dots \mathcal{D}_{M_1}
        +
        \dots.
    \end{equation}
    The  superfields $G^{++}$ introduced in this way are manifestly invariant under $\mathcal{N}=2$ supersymmetry transformations, but they are not analytic though
    can be obviously expressed through the original analytic prepotentials (see below).
    The analyticity condition \eqref{eq: an cond} leads to certain relations between $G^{++}$-superfields.
    Specifically, expanding the commutator $[\mathcal{D}^+_{\hat{\rho}}, \hat{\mathcal{H}}^{++}_s
    ]= 0$ in powers of the covariant derivatives $\mathcal{D} _M$
    and equating the coefficient of each independent derivative combination to zero yields a system of differential relations.
    As an example, we give the relations for superfields with totally symmetrized indices, which we will use in
    the sequel \footnote{We use the standard  short-hand  notation $\alpha(s):= (\alpha_1\alpha_2\dots\alpha_s) $.}:
    \begin{subequations}\label{eq: an cond G++}
                \begin{align}
            & \mathcal{D}^+_\rho G^{++\alpha(s-1)\dot{\alpha}(s-1)} + 4i (s-1) \delta_\rho^{(\alpha} G^{++\alpha(s-2)) \dot{\alpha}(s-1) +} = 0, \\
            & \mathcal{D}^+_\rho G^{++\alpha(s-1)\dot{\alpha}(s-2)+}
            -
             \delta_\rho^{(\alpha} G^{++\alpha(s-2)) \dot{\alpha} (s-2) ++}
        \\&\qquad\qquad\qquad\qquad\quad     -
             4i (s-2) \delta_\rho^{(\alpha} G^{++\alpha(s-3)\dot{\alpha}(s-3)\alpha)+ \dot{\alpha})+}
             = 0, \\
            & \bar{\mathcal{D}}^+_{\dot{\rho}} G^{++\alpha(s-1)\dot{\alpha}(s-2)+} = 0, \\
            & \mathcal{D}^+_{\hat{\rho}} G^{++\alpha(s-2)\dot{\alpha}(s-2)++} = 0.
        \end{align}
    \end{subequations}
Analogously, in the covariant basis, the operator \eqref{eq: Lambda k} is given by
\begin{equation}\label{eq: Lambda k cov}
    \hat{\Lambda}_s^{(k)}
     =
      \Lambda^{M_1\dots M_{k-1}} \mathcal{D}_{M_{k-1}} \dots \mathcal{D}_{M_1},
\end{equation}
where the  parameters $\Lambda$ satisfy analytic constraints  analogous to those for the $G^{++}$-superfields. For later use, we record the following relations:
    \begin{subequations}
    \begin{align}
        & \mathcal{D}^+_\rho \Lambda^{\alpha(s-1)\dot{\alpha}(s-1)} + 4i (s-1) \delta_\rho^{(\alpha} \Lambda^{\alpha(s-2)) \dot{\alpha}(s-1) +} = 0, \\
        & \mathcal{D}^+_\rho \Lambda^{\alpha(s-1)\dot{\alpha}(s-2)+}
    -
     \delta_\rho^{(\alpha} \Lambda^{\alpha(s-2)) \dot{\alpha} (s-2) ++}         \\&\qquad\qquad\qquad\qquad  -
        4i (s-2) \delta_\rho^{(\alpha} \Lambda^{\alpha(s-3)\dot{\alpha}(s-3)\alpha)+ \dot{\alpha})+}
        = 0,  \\
        & \bar{\mathcal{D}}^+_{\dot{\rho}} \Lambda^{\alpha(s-1)\dot{\alpha}(s-2)+} = 0, \\
        & \mathcal{D}^+_{\hat{\rho}} \Lambda^{\alpha(s-2)\dot{\alpha}(s-2)++} = 0.
    \end{align}
\end{subequations}
These relations are a direct generalization of the analogous ones in the spin-$\mathbf{2}$ case \cite{Ivanov:2025gvs}.

\medskip

The transformation law \eqref{eq: transf conv} in the covariant basis reads
\begin{equation}
    \begin{split}
    \delta^{(s,k)}_\lambda \hat{\mathcal{H}}^{++}_s
    =
    [\mathcal{D}^{++}, \hat{\Lambda}_s^{(k)}]
    =&\,
    \mathcal{D}^{++} \Lambda^{M_1\dots M_{k-1}} \mathcal{D}_{M_{k-1}} \dots \mathcal{D}_{M_1}
\\
&   +
      \Lambda^{M_1\dots M_{k-1}}  [ \mathcal{D}^{++},  \mathcal{D}_{M_{k-1}} \dots \mathcal{D}_{M_1} ].
        \end{split}
\end{equation}
The second line involves two types of non-trivial commutators,
\begin{equation}
    [\mathcal{D}^{++}, \mathcal{D}^-_{\hat{\alpha}}]
    =
    \mathcal{D}^+_{\hat{\alpha}},
    \qquad
        [\mathcal{D}^{++}, \mathcal{D}^{--}]
    =
    \mathcal{D}^0\,,
\end{equation}
which will then generate contributions with $k-3$ superspace derivatives, when being ``sandwiched'' between two analytic hypermultiplets.

Because the leading terms in  $\hat{\mathcal{H}}^{++}_s$
contain at least $s-2$ space-time derivatives, which commute with $\mathcal{D}^{++}$, the top-level $G^{++}$-superfields (with $s-1$ indices) obey a simple, universal gauge transformation law\footnote{If one uses non-covariant derivatives,
commutator terms arise:
$
        [\mathcal{D}^{++}, \partial^-_\alpha] = 4i \bar{\theta}^{+\dot{\alpha}} \partial_{\alpha\dot{\alpha}} - \partial^+_\alpha$,
    that leads to the shift contributions;  see, e.g., Eq. \eqref{eq: s=3 gauge transformations vector} in the spin-$\mathbf{3}$ case.}
\begin{equation}\label{eq: G++ superfields}
    \delta_\lambda G^{++M_1\dots M_{s-1}} = \mathcal{D}^{++} \Lambda^{M_1\dots M_{s-1}}.
\end{equation}
The lower-order superfields with fewer numbers of indices ({\it i.e.}, with $s-3$, $s-5$ and so on)
will acquire additional shift contributions, i.e. terms that do not reduce to harmonic derivatives and so have the form $\sim \Lambda^{M_1\dots M_k}$.

With the general geometric framework and the covariant  $G^{++}$-superfields having been defined, we are equipped with all means needed to construct the explicit dynamical invariants.
%In the next Section, we will exemplify this machinery by constructing an action for the spin-$\mathbf{3}$ case.

\section{$\mathcal{N}=2$ super Weyl actions}

We now turn to constructing dynamical actions for superconformal multiplets.
To illustrate the general procedure, we first consider in detail
the spin-$\mathbf{3}$ superconformal multiplet. As the first non-trivial example of a higher-spin theory ($s > 2$), the spin-$\mathbf{3}$
case serves as a prototype, exhibiting all the key structural features of the general integer-spin formulation, while avoiding an  unnecessary index clutter.

\subsection{Example: spin-$\mathbf{3}$ Weyl theory}\label{sec: spin-3 theory}

%We begin by explicitly applying the general framework of Section~\ref{sec: multiplets} to the spin-$\mathbf{3}$ multiplet.
In the spin-$\mathbf{3}$ case, the analytic differential operator \eqref{eq: sc spin s operator} takes the following explicit form:
\begin{equation}\label{eq: spin 3 operator}
    \begin{split}
    \hat{\mathcal{H}}^{++}_{(s=3)}
    =\;&
    h^{++\alpha\dot{\alpha} \beta\dot{\beta}} \partial_{\beta\dot{\beta}}  \partial_{\alpha\dot{\alpha}}
    +
        2h^{++ \beta\dot{\beta} \alpha+} \partial_{\beta\dot{\beta}} \partial^-_{\alpha}
            +
        2h^{++ \beta\dot{\beta} \dot{\alpha}+} \partial_{\beta\dot{\beta}} \partial^-_{\dot{\alpha}}
        +
        2   h^{++ \beta\dot{\beta}++}
            \partial_{\beta\dot{\beta}}
         \partial^{--}
            \\
            &
            +
            h^{++[\alpha+\beta]+} \partial^-_{\beta}
            \partial^-_{\alpha}
            +
            h^{++[\dot{\alpha}+\dot{\beta}]+} \partial^-_{\dot{\beta}}
            \partial^-_{\dot{\alpha}}
            +
        2   h^{++\alpha+\dot{\beta}+} \partial^-_{\dot{\beta}}
            \partial^-_{\alpha}
            \\
            &+
            2 h^{++\hat{\alpha}+++} \partial^{--}
                \partial^-_{\hat{\alpha}}
                +
                h^{(+6)} \partial^{--} \partial^{--} +
                h^{++},
            \end{split}
\end{equation}
where $h^{++\dots}$ are unconstrained analytic higher-spin prepotentials.
We do not assume any special symmetry of the prepotential indices.
The gauge transformations of analytic prepotentials are given by:
\begin{equation}\label{eq: s=3 gauge transformations vector}
    \begin{split}
        \delta_\lambda h^{++\alpha\dot{\alpha}\beta\dot{\beta}} &=      \mathcal{D}^{++}\lambda^{\alpha\beta\dot{\alpha}\dot{\beta}}
        + 2i \left( \lambda^{\alpha\dot{\alpha}\beta+} \bar{\theta}^{+\dot{\beta}}
        + \lambda^{\beta\dot{\beta}\alpha+} \bar{\theta}^{+\dot{\alpha}}\right)
        \\&\qquad\qquad\qquad
        - \,2i
        \left( \bar{\lambda}^{\alpha\dot{\alpha}\dot{\beta}+} \theta^{+\beta}
        + \bar{\lambda}^{\beta\dot{\beta}\dot{\alpha}+} \theta^{+\alpha}
        \right),
        \\
        \delta_\lambda h^{++\alpha\dot{\alpha}\beta+} &=  \mathcal{D}^{++}
        \lambda^{\alpha\dot{\alpha}\beta+}
        - 8i \lambda^{[\alpha+\beta]+} \bar{\theta}^{+\dot{\alpha}}
        -
        8i \lambda^{\beta+\dot{\alpha}+} \theta^{+\alpha}
        - \lambda^{\alpha\dot{\alpha}++} \theta^{+\beta},
        \\
        \delta_\lambda h^{++\alpha\dot{\alpha}++} &= \mathcal{D}^{++} \lambda^{++\alpha\dot{\alpha}}
        + 4 i \lambda^{++\alpha+} \bar{\theta}^{+\dot{\alpha}}
        -
        4i \lambda^{++\dot{\alpha}+}\theta^{+\alpha}.
    \end{split}
\end{equation}
The transformation laws of the remaining spin-$\mathbf{3}$ analytic prepotentials can be found in ref.~\cite{Buchbinder:2024pjm}.
%we will not use them here.

To ensure that this geometric construction indeed yields the $\mathcal{N}=2$ superconformal spin-$3$ multiplet,
it is instructive to examine its component content. The analysis carried out in \cite{Buchbinder:2024pjm} shows that only the symmetric part of the superfields
present in the first line of \eqref{eq: spin 3 operator} encapsulates  physical degrees of freedom in the Wess-Zumino (WZ) gauge:
\begin{equation}
    h^{++(\alpha\beta)(\dot{\alpha}\dot{\beta})},
    \quad
    h^{++(\alpha\beta)\dot{\alpha}+},
    \quad
    h^{++(\dot{\alpha}\dot{\beta})\alpha+},
    \quad
    h^{++\beta\dot{\beta}++}.
\end{equation}
Their component content in the WZ-type gauge is written as
\begin{equation}\label{eq: spin 3 GF}
        \begin{split}
            h_{WZ}^{++(\alpha\beta)(\dot{\alpha}\dot{\beta})}
            &=
            -4i \theta^{+}_{\gamma}\bar{\theta}^{+}_{\dot{\gamma}} \Phi^{(\alpha\beta\gamma)(\dot{\alpha}\dot{\beta}\dot{\gamma})}
            -
            (\bar{\theta}^+)^2 \theta^{+}_{\gamma} \psi^{(\alpha\beta\gamma)(\dot{\alpha}\dot{\beta})i} u_i^-
            \\&\;\;\;\;\;\;\;\;\;\;\;\;\;\;\;\;
            -
            (\theta^+)^2 \bar{\theta}^{+}_{\dot{\gamma}} \bar{\psi}^{(\alpha\beta)(\dot{\alpha}\dot{\beta}\dot{\gamma})i} u_i^-
            +
            (\theta^+)^2 (\bar{\theta}^+)^2 V^{(\alpha\beta)(\dot{\alpha}\dot{\beta})ij}u^-_i u^-_j\,,
            \\
            h_{WZ}^{++(\alpha\beta)\dot{\alpha}+}
            &=
            (\theta^{+})^2 \bar{\theta}^{+}_{\dot{\beta}} \mathcal{P}^{(\alpha\beta)(\dot{\alpha}\dot{\beta})}
            +
            (\bar{\theta}^{+})^2 \theta^{+}_{\gamma} T^{(\alpha\beta\gamma)\dot{\alpha}}_{}
            +
            (\theta^{+})^4 \chi^{(\alpha\beta)\dot{\alpha}i}u_i^-\,,
            \\
            h_{WZ}^{++\alpha(\dot{\alpha}\dot{\beta})+}&=
            \widetilde{h_{WZ}^{++(\alpha\beta)\dot{\alpha}+}}\,,
            \\
            h_{WZ}^{++\alpha\dot{\alpha}++} \;\,&=
            (\theta^+)^2 (\bar{\theta}^+)^2 D^{\alpha\dot{\alpha}}\,.
        \end{split}
\end{equation}
The possibility of the gauge removal of unphysical components stems directly from the algebraic structure of the gauge transformations \eqref{eq: s=3 gauge transformations vector}.
By the suitable choice of the gauge parameters, all lower-spin prepotentials ($h^{++\hat{\alpha}+++}$, $h^{(+6)}$, etc.), as well as the trace parts of the leading superfields,
can be completely gauged away.

 Physically, all component fields in \eqref{eq: spin 3 GF} are conformal higher-spin gauge fields of the Fradkin-Tseytlin type, with the residual gauge freedom acting on them
 in the standard way, e.g.\footnote{Hereafter, the complete symmetrization over the  indices of the same type is assumed in the right-hand side, if it holds for the left-hand side.},
 \begin{equation}
 \delta \Phi^{\alpha(3)\dot{\alpha}(3)} = \partial^{\alpha\dot{\alpha}} a^{\alpha(2) \dot{\alpha}(2)},
 \qquad
 \delta \psi^{\alpha(3)\dot{\alpha}(2)i}
 =
 \partial^{\alpha\dot{\alpha}} \epsilon^{\alpha(2)\dot{\alpha}i}.
 \end{equation}
  Specifically, the lowest component $\Phi^{\alpha(3)\dot{\alpha}(3)}$
 represents the bosonic conformal spin-3 gauge field (the higher-spin analogue of the Weyl graviton), while $\psi^{\alpha(3)\dot{\alpha}(2)i}$ and  $\bar{\psi}^{\alpha(2)\dot{\alpha}(3)i}$
 correspond to its fermionic superpartners (conformal spin-$5/2$ gravitini).
 The remaining components $$    V^{\alpha(2)\dot{\alpha}(2)(ij)}, \;\;\mathcal{P}^{\alpha(2)\dot{\alpha}(2)} \;\;,
 T^{\alpha(3)\dot{\alpha}},\;\;
 \chi^{\alpha(2)\dot{\alpha}i}, \;\;
 D^{\alpha\dot{\alpha}}$$ account for the lower-spin conformal gauge fields (spins 2, 3/2, 1) needed to close the supermultiplet.
The structure of the spin-$\mathbf{3}$ supermultiplet can be depicted as
\begin{equation*}
      \Phi^{\alpha(3)\dot{\alpha}(3)}
    \; \leftrightarrow \;
    \psi^{\alpha(3)\dot{\alpha}(2)i}
    \; \leftrightarrow \;
    \big((V^{\alpha(2)\dot{\alpha}(2)(ij)}, \mathcal{P}^{\alpha(2)\dot{\alpha}(2)}),
    T^{\alpha(3)\dot{\alpha}}\big)
    \;\; \leftrightarrow \;\;
    \chi^{\alpha(2)\dot{\alpha}i}
    \;\; \leftrightarrow \;\;
D^{\alpha\dot{\alpha}}\,,
\end{equation*}
    \begin{equation*}
    {\bf 3}
    \;\; \leftrightarrow \;\;
    2 \times {\bf 5/2}
    \;\; \leftrightarrow \;\;
    \big(4 \times {\bf 2}, \text{hook field (conformal pseudo-spin {\bf 2} field)}\big)
    \;\; \leftrightarrow \;\;
    2 \times {\bf 3/2}
    \;\; \leftrightarrow \;\;
    {\bf 1}.
 \end{equation*}

Thus the harmonic superspace approach naturally packages
the $\mathcal{N}=2$ superconformal spin ${\bf 3}$ gauge supermultiplet into few unconstrained analytic prepotentials.

 \subsection*{Supersymmetry-invariant superfields and super Weyl action}

In order to construct manifestly supersymmetric invariants, it is necessary to express our geometric quantities in terms of objects that transform homogeneously
under rigid $\mathcal{N}=2$ supersymmetry. It requires passing to the covariant derivative basis. In this basis,
an expansion of the differential operator \eqref{eq: spin 3 operator} is given by:
 \begin{equation}\label{eq: spin 3 operator cov}
    \begin{split}
        \hat{\mathcal{H}}^{++}_{(s=3)}
        =\;&
        G^{++\alpha\dot{\alpha} \beta\dot{\beta}} \partial_{\beta\dot{\beta}}  \partial_{\alpha\dot{\alpha}}
        +
        2G^{++\hat{\alpha}+ \beta\dot{\beta}} \partial_{\beta\dot{\beta}} \mathcal{D}^-_{\hat{\alpha}}
        +
        2G^{++ \beta\dot{\beta}++}
        \partial_{\beta\dot{\beta}}
        \mathcal{D}^{--}
        \\
    &
    +
    G^{++[\alpha+\beta]+} \mathcal{D}^-_{\beta}
    \mathcal{D}^-_{\alpha}
    +
    G^{++[\dot{\alpha}+\dot{\beta}]+} \bar{\mathcal{D}}^-_{\dot{\beta}}
    \bar{\mathcal{D}}^-_{\dot{\alpha}}
    +
    2G^{++\alpha+\dot{\beta}+} \bar{\mathcal{D}}^-_{\dot{\beta}}
    \mathcal{D}^-_{\alpha}
    \\
    &
        +
        2G^{++\hat{\alpha}+++}  \mathcal{D}^{--}
            \mathcal{D}^-_{\hat{\alpha}}
        +
        G^{(+6)}    \mathcal{D}^{--}    \mathcal{D}^{--}
         +
        G^{++}.
    \end{split}
 \end{equation}
Recall once more that the $G^{++}$-superfields appearing here are manifestly invariant under rigid $\mathcal{N}=2$ supersymmetry.
 However, they are no longer analytic. This loss of manifest analyticity occurs due to non-vanishing (anti)commutators
 of the covariant derivatives $\mathcal{D}_M$ with the spinor derivative $\mathcal{D}^+_{\hat{\rho}}$, unlike the  derivatives $\partial_M$. The analyticity condition \eqref{eq: an cond G++}
 translates into a set of non-trivial differential constraints relating different $G^{++}$-superfields. Eqs. \eqref{eq: spin 3 operator} and \eqref{eq: spin 3 operator cov}
 uniquely determine the relation between the $G^{++}$-superfields and the analytic prepotentials $h^{++}$, e.g.,
 \begin{equation}
    \begin{split}
    G^{++(\alpha\beta)(\dot{\alpha}\dot{\beta})}
    =\;&
    h^{++(\alpha\beta)(\dot{\alpha}\dot{\beta})}
    +
    4i h^{++(\alpha\beta)(\dot{\alpha}+} \bar{\theta}^{-\dot{\beta})}
    +
    4i
    \theta^{-(\alpha} h^{++\beta)(\dot{\alpha}\dot{\beta})+}
    \\&
        -
    4i h^{++(\alpha(\dot{\alpha}++}
    \theta^{-\beta)} \bar{\theta}^{-\dot{\beta})}
    -
    16 h^{++(\alpha+(\dot{\alpha}+} \theta^{-\beta)} \bar{\theta}^{-\dot{\beta})}.
    \end{split}
 \end{equation}
According to the general gauge transformation law \eqref{eq: G++ superfields}, the gauge transformation of the $G^{++}$-superfields with  symmetrized indices are given by
 \begin{equation}
    \begin{split}
    \delta_\lambda G^{++(\alpha\beta)(\dot{\alpha}\dot{\beta})}
    &=
    \mathcal{D}^{++} \Lambda^{(\alpha\beta)(\dot{\alpha}\dot{\beta})},
    \\
    \delta_\lambda
    G^{++(\alpha\beta)\dot{\alpha}+}
    &=
    \mathcal{D}^{++}
    \Lambda^{(\alpha\beta)\dot{\alpha}+},
    \\
    \delta_\lambda G^{++\alpha\dot{\alpha} ++}
    &=
    \mathcal{D}^{++} \Lambda^{\alpha\dot{\alpha} ++}.
    \end{split}
 \end{equation}

With the invariant superfields at hand, we can proceed to constructing gauge-invariant objects.
The key idea is to find specific combinations of the superspace derivatives which, when acting on the $G^{++}$-superfields, systematically strip away the dotted indices, leaving only the undotted indices on a superfield, still preserving some modified analyticity.
 Furthermore, the variation of such specific combinations must vanish under the chiral projector $\left( \bar{\mathcal{D}}^+ \right)^2$.
 As a generalization of the spin-$\mathbf{2}$ half-analytic superfield of Ref. \cite{Ivanov:2024gjo, Ivanov:2025gvs, Ivanov:2025mld}, we introduce
\begin{equation}\label{eq: spin-3 HA}
    \mathcal{H}^{++}_{(\alpha_1\alpha_2\alpha_3\alpha_4)}
    :=
    \partial_{(\alpha_1}^{\dot{\beta}_1}
    \partial_{\alpha_2}^{\dot{\beta}_2}
    G^{++}_{\alpha_3\alpha_4) (\dot{\beta}_1\dot{\beta}_2)}
    +
    2\partial_{(\alpha_1}^{\dot{\beta}}
    \mathcal{D}^-_{\alpha_2}  G^{+++}_{\alpha_3\alpha_4)\dot{\beta}}.
\end{equation}
This superfield satisfies the crucial \textit{half-analyticity condition}
\begin{equation}
    \bar{\mathcal{D}}^+_{\dot{\rho}}    \mathcal{H}^{++}_{\alpha(4)}
    =
    0.
\end{equation}
It can be directly   checked by using the relations \eqref{eq: an cond G++} for $s=3$ case and the algebra of covariant derivatives \eqref{eq: rel cov derivatives}.

The gauge transformations \eqref{eq: s=3 gauge transformations vector} acts on the half-analytic superfield \eqref{eq: spin-3 HA} as
\begin{equation}\label{eq: spin 3 gauge}
    \delta_\lambda
    \mathcal{H}^{++}_{\alpha(4)}
    =
    \mathcal{D}^{++} \Lambda_{\alpha(4)},
\end{equation}
where the  gauge superparameter equals
\begin{equation}
    \Lambda^{}_{(\alpha_1\alpha_2\alpha_3\alpha_4)}
    =
    \partial_{(\alpha_1}^{\dot{\beta}_1}
    \partial_{\alpha_2}^{\dot{\beta}_2}
    \Lambda^{}_{\alpha_3\alpha_4) (\dot{\beta}_1\dot{\beta}_2)}
    +
    2\partial_{(\alpha_1}^{\dot{\beta}}
    \mathcal{D}^-_{\alpha_2}  \Lambda^{+}_{\alpha_3\alpha_4)\dot{\beta}}
\end{equation}
and also satisfies the half-analyticity condition,
\begin{equation}
    \bar{\mathcal{D}}^+_{\dot{\rho}}    \Lambda_{\alpha(4)}
=
0.
\end{equation}

Before turning to the construction of the gauge-invariant super-Weyl tensor, we introduce the negatively charged counterpart $\mathcal{H}^{--}_{\alpha(4)}$
of the superfield \eqref{eq: spin-3 HA}.
Following the standard harmonic superspace routine, $\mathcal{H}^{--}_{\alpha(4)}$ is defined as a  unique solution to
the harmonic zero-curvature (or ``harmonic flatness'') condition:
\begin{equation}\label{eq: zero curvature}
    \mathcal{D}^{++} \mathcal{H}^{--}_{\alpha(4)} = \mathcal{D}^{--} \mathcal{H}^{++}_{\alpha(4)}.
\end{equation}
Performing the gauge variation of this equation and utilizing the commutation relation $[\mathcal{D}^{++}, \mathcal{D}^{--}] = \mathcal{D}^0$,
one readily deduces the gauge transformation law of the newly defined superfield
%for the newly introduced negatively charged superfield:
\begin{equation}
    \delta_\lambda \mathcal{H}^{--}_{\alpha(4)} = \mathcal{D}^{--} \Lambda_{\alpha(4)}.
\end{equation}

With this object at hand, we can construct the sought gauge-invariant spin-$\mathbf{3}$ super Weyl tensor:
\begin{equation}\label{eq: spin 3 Weyl tensor}
    \mathcal{W}_{\alpha(4)}
    :=
    (\bar{\mathcal{D}}^+)^2     \mathcal{H}^{--}_{\alpha(4)},
\qquad
    \delta_\lambda \mathcal{W}_{\alpha(4)}
    =
    0.
\end{equation}
The gauge invariance $\delta_\lambda \mathcal{W}_{\alpha(4)} = 0$ is valid in a non-trivial way: the chiral projector $(\bar{\mathcal{D}}^+)^2$
hits the gauge variation $\mathcal{D}^{--}\Lambda_{\alpha(4)}$ and annihilates it precisely because $\Lambda_{\alpha(4)}$ is half-analytic,
$\bar{\mathcal{D}}^+_{\dot{\rho}} \Lambda_{\alpha(4)} = 0$.
Furthermore, applying the chiral projector $\left( \bar{\mathcal{D}}^+ \right)^2$
to the superfield $\mathcal{H}^{--}_{\alpha(4)}$
automatically guarantees its both chirality and harmonic independence:
\begin{equation}
    \mathcal{D}^{\pm\pm} \mathcal{W}_{\alpha(4)}  = 0,
    \qquad
    \bar{\mathcal{D}}^\pm_{\dot{\alpha}}
    \mathcal{W}_{\alpha(4)}  = 0.
\end{equation}
The spin-$\mathbf{3}$ super Weyl tensor contains the component spin-$\mathbf{3}$ Weyl tensor,
\begin{equation}
        \mathcal{W}_{\alpha(4)}
        \to
     \theta^{+(\alpha} \theta^{-\alpha)}C_{\alpha(6)},
     \qquad
     C_{\alpha(6)} = (\partial^3)_{\alpha(3)}^{\dot{\beta}(3)} \Phi_{\alpha(3)\dot{\beta}(3)}.
\end{equation}
This confirms that $\mathcal{W}_{\alpha(4)}   $ is indeed $\mathcal{N}=2$ supersymmetric counterpart of the linearized conformal spin-$3$ Weyl tensor.

\smallskip

Now the gauge-invariant action can be constructed similarly to the spin-$\mathbf{2}$ case as
\begin{equation}\label{eq: spin 3 action}
    \begin{split}
 S^{s=3}_{\text{Weyl}}
 =&
 \int d^4xd^8\theta du\; \mathcal{H}^{++\alpha(4)} \mathcal{H}^{--}_{\alpha(4)}
 +
 c.c.
 \\=&
 \frac{1}{16}
 \int d^4x_L d^4\theta\; \mathcal{W}^{\alpha(4)}
 \mathcal{W}_{\alpha(4)}
 +
 c.c.\,.
    \end{split}
\end{equation}
Note that the imaginary counterpart of \eqref{eq: spin 3 action} yields a total derivative under the integral.

The gauge invariance of \eqref{eq: spin 3 action} in the full harmonic superspace can be directly checked:
\begin{equation}
    \begin{split}
    \delta_{\lambda} S^{s=3}_{\text{Weyl}}
    =&
\,2 \int d^4xd^8\theta du\; \mathcal{D}^{--} \Lambda^{\alpha(4)} \mathcal{H}^{++}_{\alpha(4)}
\\
=&
\,\frac{1}{128} \int d^4x du (\mathcal{D}^+)^2 (\mathcal{D}^-)^2 (\bar{\mathcal{D}}^-)^2 \;  \Big[ \left\{ (\bar{\mathcal{D}}^+)^2 \mathcal{D}^{--} \Lambda^{\alpha(4)} \right\} \mathcal{H}^{++}_{\alpha(4)} \Big] = 0.
    \end{split}
\end{equation}

The gauge invariance in the chiral representation is manifest. The chiral representation (2nd line  in \eqref{eq: spin 3 action})
is obtained from the analytic one using the following  relation in the central basis
\begin{equation}
        \mathcal{W}_{\alpha(4)}
    :=
    (\bar{\mathcal{D}}^+)^2     \mathcal{H}^{--}_{\alpha(4)}
    =
    \int du_1 (\mathcal{D}^+)^2 \frac{\mathcal{H}^{++}_{\alpha(4)}}{(u^+u_1^+)^2}
    =
    \int du (\mathcal{D}^-)^2 \mathcal{H}^{++}_{\alpha(4)},
\end{equation}
whence:
\begin{equation}
    \begin{split}
     \int d^4xd^8\theta du\; \mathcal{H}^{++\alpha(4)} \mathcal{H}^{--}_{\alpha(4)}
     =&
     \frac{1}{16}
     \int d^4x_L d^4\theta\; \left( \bar{\mathcal{D}}^-\right)^2 \mathcal{H}^{++\alpha(4)}
     \left( \bar{\mathcal{D}}^+ \right)^2
     \mathcal{H}^{--}_{\alpha(4)}
\\   =&
      \frac{1}{16}
     \int d^4x_L d^4\theta\; \mathcal{W}^{\alpha(4)}
     \mathcal{W}_{\alpha(4)}.
     \end{split}
\end{equation}

\subsection{Generalization to arbitrary integer spin $\mathbf{s}$}\label{sec: spin s theory}

The above $s=3$ construction provides a universal recipe of setting up the superconformal action for any integer spin $s$:
one isolates all $s-1$ dotted indices of the leading covariant superfield $G^{++}_{\alpha(s-1)\dot{\beta}(s-1)}$ by contracting it with $s-1$ space-time derivatives,
then forms a half-analytic superfield $\mathcal{H}^{++}_{\alpha(2s-2)}$, passes to its negatively charged counterpart via the proper harmonic flatness
conditions and, finally, projects out the relevant generalized Weyl tensor.

To be more precise, we introduce the composite superfield
\begin{equation}\label{eq: mathcal H spin s}
    \mathcal{H}^{++}_{\alpha(2s-2)}
    :=
    \left(\partial^{s-1}\right)_{\alpha(s-1)}^{\dot{\beta}(s-1)} G^{++}_{\alpha(s-1)\dot{\beta}(s-1)}
    +
    (s-1)
    \left(\partial^{s-2}\right)_{\alpha(s-2)}^{\dot{\beta}(s-2)}
    \mathcal{D}^-_\alpha G^{+++}_{\alpha(s-1)\dot{\beta}(s-2)},
\end{equation}
which satisfies the half-analyticity condition
\begin{equation}
 \bar{\mathcal{D}}^+_{\dot{\rho}}   \mathcal{H}^{++}_{\alpha(2s-2)}
 =
 0.
\end{equation}
The numerical coefficient $(s-1)$ in the second term is uniquely fixed: it originates from the expansion of the differential operator $\hat{\mathcal{H}}^{++}_s$ in the covariant basis.
Just this precise value of the relative coefficient ensures this superfield to satisfy the half-analyticity condition for any integer spin $s$.

The gauge transformation law of the superfield \eqref{eq: mathcal H spin s} has the universal form
\begin{equation}\label{eq: H++ gauge}
    \delta_\lambda \mathcal{H}^{++}_{\alpha(2s-2)}
    =
    \mathcal{D}^{++}\Lambda_{\alpha(2s-2)}\,,
\end{equation}
with the half-analytic gauge parameter
\begin{equation}
    \Lambda_{\alpha(2s-2)}
    =
    \left(\partial^{s-1}\right)_{\alpha(s-1)}^{\dot{\beta}(s-1)} \Lambda_{\alpha(s-1)\dot{\beta}(s-1)}
    +
    (s-1)
    \left( \partial^{s-2}\right)^{\dot{\beta}(s-2)}_{\alpha(s-2)} \mathcal{D}^-_{\alpha} \Lambda^+_{\alpha(s-1)\dot{\beta}(s-2)}.
\end{equation}
We can construct the appropriate negatively charged superfields by making use of the zero-curvature condition,
\begin{equation}
    \mathcal{D}^{++}  \mathcal{H}^{--}_{\alpha(2s-2)}
    =
        \mathcal{D}^{--}  \mathcal{H}^{++}_{\alpha(2s-2)}\,, \quad \delta_\lambda \mathcal{H}^{--}_{\alpha(2s-2)}
    =
    \mathcal{D}^{--}\Lambda_{\alpha(2s-2)}\,.
\end{equation}
From these superfields we can compose the gauge-invariant half-analytic spin-$\mathbf{s}$ super Weyl tensors,
\begin{equation}
    \mathcal{W}_{\alpha(2s-2)} := \left( \bar{\mathcal{D}}^+ \right)^2      \mathcal{H}^{--}_{\alpha(2s-2)},
    \qquad \bar{\mathcal{D}}^+_{\dot{\rho}}\mathcal{W}_{\alpha(2s-2)} =0\,, \;
    \delta_\lambda \mathcal{W}_{\alpha(2s-2)} = 0.
\end{equation}
As a consequence of its half-analyticity, the super Weyl  tensor satisfies the chirality and harmonic-independence conditions:
\begin{equation}
    \bar{\mathcal{D}}^\pm_{\dot{\rho}}  \mathcal{W}_{\alpha(2s-2)} = 0,
    \qquad
    \mathcal{D}^{\pm\pm} \mathcal{W}_{\alpha(2s-2)} = 0.
\end{equation}

Finally, the gauge-invariant superconformal action for an arbitrary integer spin $\mathbf{s}$ is written as:
\begin{equation}\label{eq: spin s action}
    \begin{split}
        S^{}_{\text{Weyl}}
        =&
        \int d^4xd^8\theta du\; \mathcal{H}^{++\alpha(2s-2)} \mathcal{H}^{--}_{\alpha(2s-2)}
        +
        c.c.
        \\=&\,
        \frac{1}{16}
        \int d^4x_L d^4\theta\; \mathcal{W}^{\alpha(2s-2)}
        \mathcal{W}_{\alpha(2s-2)}
        +
        c.c.
    \end{split}
\end{equation}
The passing from the representation in the full harmonic superspace to the chiral representation is performed in the same way
as in the spin-$\mathbf{3}$ case. Gauge invariance can also be checked analogously, and it is ensured by the half-analyticity
of the composite superfield \eqref{eq: mathcal H spin s}.

Thus we have successfully constructed the exact superconformal dynamical linearized actions for an arbitrary integer spin.
The final crucial step is to rigorously prove the invariance of these actions under the full rigid $\mathcal{N}=2$ superconformal group.
We address this issue in the next Section.

\section{$\mathcal{N}=2$ superconformal invariance}\label{sec: superconformal transformations}

The realization of $\mathcal{N}=2$ superconformal group in the harmonic superspace is rather non-standard \cite{Galperin:1985zv} and this entails some difficulties
in checking superconformal invariance (see, e.g., \cite{Ivanov:2005qf, Buchbinder:2020yvf}).
%In particular, the most natural method is to verify invariance in the conformal supersymmetry sector
%(or in the sector of special conformal transformations),
In a recent work \cite{Ivanov:2025gvs}, we used some universal method for checking superconformal invariance of the harmonic
action of the  linearized Weyl supergravity.
Its key advantage is that various relations between the conformal superparameters allow one to obtain a simple superfield form of the transformation
 law. In this Section, we apply this method to check superconformal invariance of the actions
of conformal higher spins \eqref{eq: spin 3 action} and \eqref{eq: spin s action}.

\subsection{$\mathcal{N}=2$ superconformal group in harmonic superspace}\label{eq: sec rel Lambda}

We start by listing some useful relations between the superconformal parameters.

Recall first that rigid $\mathcal{N}=2$ superconformal transformations in HSS are described by the first-order differential operator
\begin{equation}\label{eq: Lambda operator}
    \hat{\Lambda} : = \lambda^{\alpha\dot{\alpha}}
    \partial_{\alpha\dot{\alpha}}
    +
    \lambda^{+\hat{\alpha}} \partial^-_{\hat{\alpha}}
    +
        \lambda^{-\hat{\alpha}} \partial^+_{\hat{\alpha}}
        +
        \lambda^{++}\partial^{--}\,,
\end{equation}
which satisfies the relation
\begin{equation}\label{eq: N=2 sc}
    \delta_{sc}\mathcal{D}^{++} := [ \hat{\Lambda}_{sc}, \mathcal{D}^{++}] = -  \lambda^{++}_{sc} \mathcal{D}^0\,.
\end{equation}
This condition amounts to a set of equations for the superconformal parameters:
\begin{equation}\label{eq: sc equations}
    \begin{cases}
        \mathcal{D}^{++} \lambda_{sc}^{\alpha\dot{\alpha}}
        +
        4i \lambda_{sc}^{+\alpha} \bar{\theta}^{+\dot{\alpha}}
        +
        4i \theta^{+\alpha} \bar{\lambda}_{sc}^{+\dot{\alpha}} = 0,
        \\
        \mathcal{D}^{++} \lambda_{sc}^{+\hat{\alpha}} - \theta^{+\hat{\alpha}} \lambda_{sc}^{++} = 0,
        \\
        \mathcal{D}^{++} \lambda_{sc}^{-\hat{\alpha}} - \lambda_{sc}^{+\hat{\alpha}}
        + \theta^{-\hat{\alpha}} \lambda_{sc}^{++}  = 0,
        \\
        \mathcal{D}^{++} \lambda_{sc}^{++} = 0.
    \end{cases}
\end{equation}
The most general solution of eqs. \eqref{eq: sc equations} reads \cite{Buchbinder:2024pjm}
\footnote{We have skipped here the translations, Lorentz, and global $\mathcal{N}=2$ supersymmetry transformations, since
the invariance of the superfield actions under these transformations is manifest.}:
\begin{equation}\label{eq:superconformal symmetry}
    \begin{split}
        %   \begin{cases}
            \lambda_{sc}^{\alpha\dot{\alpha}}
            =\;&
            x^{\dot{\alpha}\beta} k_{\beta\dot{\beta}} x^{\dot{\beta}\alpha}
            %   \\&
            +
            {\rm d} x^{\alpha\dot{\alpha}}
            -
            4i \theta^{+\alpha} \bar{\theta}^{+\dot{\alpha}} \lambda^{(ij)}u^-_i u^-_j
            -
            4i \left(x^{\alpha\dot{\beta}}\bar{\eta}_{\dot{\beta}}^i \bar{\theta}^{+\dot{\alpha}}
            +
            \theta^{+\alpha} \eta^i_{\beta} x^{\beta\dot{\alpha}}
            \right) u^-_i,
            \\
            \lambda^{+\alpha}_{sc}
            =\;&
            \frac{1}{2}
            ({\rm d} + i {\rm r})\theta^{+\alpha}
            +
            x^{\alpha\dot{\beta}} k_{\beta\dot{\beta}} \theta^{+\beta}
            +
            x^{\alpha\dot{\beta}}  \bar{\eta}^i_{\dot{\beta}} u^+_i
            %   \\&
            +
            \theta^{+\alpha}
            \left( \lambda^{(ij)}u^+_i u^-_j
            +
            4i \theta^{+\beta} \eta^i_{\beta} u^-_i
            \right),
            \\
            \bar{\lambda}^{+\dot{\alpha}}_{sc}
            =\;&
            \frac{1}{2}
            ({\rm d} - i {\rm r})\bar{\theta}^{+\dot{\alpha}}
            +
            x^{\dot{\alpha}\beta} k_{\beta\dot{\beta}} \bar{\theta}^{+\dot{\beta}}
            +
            x^{\beta\dot{\alpha}}  \eta^i_{\beta} u^+_i
            %   \\&
            +
            \bar{\theta}^{+\dot{\alpha}}
            \left( \lambda^{(ij)}u^+_i u^-_j
            -
            4i \bar{\theta}^{+\dot{\beta}} \bar{\eta}^i_{\dot{\beta}} u^-_i
            \right),
            \\
            \lambda^{-\alpha}_{sc} =\;&
            \frac{1}{2} ({\rm d}+i{\rm r})\theta^{-\alpha}
            +
            x^{\alpha\dot{\beta}}
            k_{\beta\dot{\beta}}   \theta^{-\beta}
            -
            2i (\theta^-)^2 \bar{\theta}^+_{\dot{\beta}} k^{\dot{\beta}\alpha}
            + \left( x^{\alpha\dot{\beta}} + 4i \theta^{-\alpha}\bar{\theta}^{+\dot{\beta}} \right) \bar{\eta}^i_{\dot{\beta}} u^-_i
            \\& +
            4i \eta^i_{\beta} \theta^{-\beta} \left( \theta^{-\alpha} u^+_i - \theta^{+\alpha} u^-_i \right)
            +
            \lambda^{ij} u^-_i \left( u^-_j \theta^{+\alpha} - u^+_j \theta^{-\alpha} \right),
            \\
            \bar{\lambda}^{-\dot{\alpha}}_{sc}
            =\;&
            \frac{1}{2}({\rm d}-i{\rm r}) \bar{\theta}^{-\dot{\alpha}}
            +
            x^{\dot{\alpha}\beta} k_{\beta\dot{\beta}} \bar{\theta}^{-\dot{\beta}}
            -
            2i (\bar{\theta}^-)^2 \theta^+_{\beta} k^{\beta\dot{\alpha}}
            +
            \left( x^{\beta\dot{\alpha}} + 4i \theta^{+\beta} \bar{\theta}^{-\dot{\alpha}} \right) \eta^i_{\beta} u^-_i
            \\&
            -
            4i \bar{\eta}^i_{\dot{\beta}} \bar{\theta}^{-\dot{\beta}} \left( \bar{\theta}^{-\dot{\alpha}} u^+_i - \bar{\theta}^{+\dot{\alpha}} u^-_i \right)
            +
            \lambda^{ij} u^-_i \left( u^-_j \bar{\theta}^{+\dot{\alpha}} - u^+_j \bar{\theta}^{-\dot{\alpha}} \right),
            \\
            \lambda^{++}_{sc} =\;&
            \lambda^{ij} u^+_i u^+_j
            +
            4i \theta^{+\beta} \bar{\theta}^{+\dot{\beta}} k_{\beta\dot{\beta}}
            +
            4i \left( \theta^{+\beta} \eta^i_{\beta} + \bar{\eta}^i_{\dot{\beta}} \bar{\theta}^{+\dot{\beta}}  \right) u^+_i.
            %   \end{cases}
    \end{split}
\end{equation}
The parameters appearing here correspond to special conformal symmetry ($k_{\alpha\dot{\alpha}}$), dilatations (${\rm d}$), $U(1)_R$ rotations (${\rm r}$), conformal supersymmetry ($\eta^i_\alpha$), and $SU(2)_R$ rotations ($\lambda^{(ij)}$).

\medskip

In what follows, it will be convenient  to rewrite the operator \eqref{eq: Lambda operator} in the basis of covariant derivatives,
\begin{equation}\label{eq: Lambda operator cov}
    \hat{\Lambda} : = \Lambda^{\alpha\dot{\alpha}}
    \partial_{\alpha\dot{\alpha}}
    +
    \Lambda^{+\hat{\alpha}} \mathcal{D}^-_{\hat{\alpha}}
    +
    \Lambda^{-\hat{\alpha}} \mathcal{D}^+_{\hat{\alpha}}
    +
    \Lambda^{++}\mathcal{D}^{--}.
\end{equation}
This will allow us to obtain a set of useful relations in a simple way.

The condition \eqref{eq: N=2 sc} can be rewritten as
\begin{equation}\label{eq: N=2 sc 1}
    \delta_{sc}\mathcal{D}^{++} := [ \hat{\Lambda}_{sc}, \mathcal{D}^{++}] = -  \Lambda^{++}_{sc} \mathcal{D}^0\,,
\end{equation}
and it is equivalent to a system of equations for the $\Lambda_{sc}$ parameters:
\begin{equation}
    \begin{cases}
        \mathcal{D}^{++}    \Lambda_{sc}^{\alpha\dot{\alpha}} = 0,
        \\
        \mathcal{D}^{++}    \Lambda_{sc}^{+\hat{\alpha}} = 0,
        \\
        \mathcal{D}^{++}    \Lambda_{sc}^{-\hat{\alpha}}
        +
        \Lambda_{sc}^{+\hat{\alpha}}
        =
        0,
        \\
        \mathcal{D}^{++} \Lambda_{sc}^{++} = 0.
    \end{cases}
\end{equation}
Likewise, the transformation law for the negatively charged harmonic derivative,
\begin{equation}\label{eq: N=2 sc 2}
    \delta_{sc} \mathcal{D}^{--} := [\hat{\Lambda}_{sc}, \mathcal{D}^{--}]  =  - (\mathcal{D}^{--}\Lambda_{sc}^{++}) \mathcal{D}^{--}\,,
\end{equation}
amounts to the set of equations:
\begin{equation}\label{eq: d-- rel}
    \begin{cases}
        \mathcal{D}^{--}    \Lambda_{sc}^{\alpha\dot{\alpha}} = 0,
        \\
        \mathcal{D}^{--}    \Lambda_{sc}^{+\hat{\alpha}}
        +
        \Lambda_{sc}^{-\hat{\alpha}}  = 0,
        \\
        \mathcal{D}^{--}    \Lambda_{sc}^{-\hat{\alpha}}
        =
        0.
    \end{cases}
\end{equation}
From these relations some important identities for $\Lambda_{sc}^{\alpha\dot{\alpha}}$ follow:
\begin{subequations}
    \begin{equation}\label{eq: D-beta}
        \mathcal{D}^-_{\beta} \Lambda_{sc}^{\alpha\dot{\alpha}} =
        4i \delta_\beta^\alpha \bar{\Lambda}_{sc}^{-\dot{\alpha}},
    \qquad
        \bar{\mathcal{D}}^-_{\dot{\beta}} \Lambda_{sc}^{\alpha\dot{\alpha}}
        =
        -
        4i \delta^{\dot{\alpha}}_{\dot{\beta}} \Lambda_{sc}^{-\alpha},
    \end{equation}
    \begin{equation}\label{eq: partial alpha dot alpha}
        \partial_{\beta\dot{\beta}} \Lambda_{sc}^{\alpha\dot{\alpha}}
        =
        \delta^\alpha_\beta \left( \bar{\mathcal{D}}^-_{\dot{\beta}}  \bar{\Lambda}_{sc}^{+\dot{\alpha}}\right)
        +
        \delta^{\dot{\alpha}}_{\dot{\beta}} \left( \mathcal{D}^-_\beta \Lambda_{sc}^{+\alpha} \right) - \delta^\alpha_\beta \delta^{\dot{\alpha}}_{\dot{\beta}} \left( \mathcal{D}^{--} \Lambda_{sc}^{++}\right).
    \end{equation}
Applying the derivatives $\mathcal{D}^+_\alpha$ and $\bar{\mathcal{D}}^+_{\dot{\alpha}}$ to eq. \eqref{eq: partial alpha dot alpha} also yields:
    \begin{equation}
        4i\partial_{\beta\dot{\beta}} \Lambda^{+\alpha}_{sc} =- \delta_\beta^\alpha \bar{\mathcal{D}}^-_{\dot{\beta}} \Lambda^{++}_{sc},
    \qquad
        4i \partial_{\beta\dot{\beta}} \bar{\Lambda}^{+\dot{\alpha}}_{sc} = \delta^{\dot{\alpha}}_{\dot{\beta}} \mathcal{D}^-_{\beta} \Lambda^{++}_{sc}   .
    \end{equation}
\end{subequations}

The superconformal transformations of higher-spin superfields involve the matrix $\mathcal{D}_N \Lambda_{sc}^M$ where  $N,M \in \{\alpha\dot{\alpha}, \hat{\alpha}+, ++ \}$.
For our purpose it will be enough to know the schematic  structure of some its entries:
\begin{subequations}
\begin{equation}
    \partial_{\alpha\dot{\alpha}} \Lambda_{sc}^{\beta\dot{\beta}} \sim (\delta_\alpha^\beta, \delta_{\dot{\alpha}}^{\dot{\beta}}),
    \qquad
    \mathcal{D}^-_{\hat{\alpha}} \Lambda_{sc}^{\beta\dot{\beta}} \sim \delta_{\hat{\alpha}}^{\hat{\beta}},
    \qquad
    \mathcal{D}^{--} \Lambda_{sc}^{\beta\dot{\beta}} = 0,
\end{equation}
\begin{equation}
    \partial_{\alpha\dot{\alpha}} \Lambda_{sc}^{+\hat{\beta}} \sim \delta_{\hat{\alpha}}^{\hat{\beta}},
    \qquad\qquad
    \mathcal{D}^-_{\hat{\alpha}} \Lambda_{sc}^{+\hat{\beta}} \sim \delta_{\hat{\alpha}}^{\hat{\beta}} ,
    \qquad
    \mathcal{D}^{--} \Lambda_{sc}^{+\hat{\beta}} \neq0,
\end{equation}
\begin{equation}
    \partial_{\alpha\dot{\alpha}} \Lambda_{sc}^{++} = 0,
    \qquad\qquad
    \mathcal{D}^-_{\hat{\alpha}} \Lambda_{sc}^{++} \neq 0,
    \qquad
        \mathcal{D}^{--} \Lambda_{sc}^{++} \neq 0.
\end{equation}
\end{subequations}
We will also use the relations,
\begin{subequations}\label{eq: Lor factors}
    \begin{equation}
        \mathcal{D}^-_\beta \Lambda_{sc}^{+\alpha}
        -
        \mathcal{D}^+_\beta \Lambda_{sc}^{-\alpha}
        =
        \delta_\beta^\alpha \left(\mathcal{D}^{--} \Lambda_{sc}^{++}\right),
    \end{equation}
    \begin{equation}
        \bar{\mathcal{D}}^-_{\dot{\beta}} \bar{\Lambda}_{sc}^{+\dot{\alpha}}
        -
        \bar{\mathcal{D}}^+_{\dot{\beta}} \bar{\Lambda}_{sc}^{-\dot{\alpha}}
        =
        \delta_{\dot{\beta}}^{\dot{\alpha}} \left(\mathcal{D}^{--} \Lambda_{sc}^{++}\right).
    \end{equation}
\end{subequations}

Finally, we will need the superconformal transformation laws of the covariant derivatives
\begin{subequations}
\begin{equation}\label{eq: D+ sc}
    \begin{split}
        & \delta_{sc} \mathcal{D}^+_{\alpha}
        :=
        [\hat{\Lambda}_{sc},  \mathcal{D}^+_{\alpha} ]
        =
        -
        \left( \mathcal{D}^+_{\alpha} \Lambda_{sc}^{-\beta} \right) \mathcal{D}^+_{\beta}\,,
    \end{split}
\end{equation}
\begin{equation}\label{eq: D- sc}
    \begin{split}
        \delta_{sc} \mathcal{D}^-_\alpha =&  -
        (\mathcal{D}^{--}\Lambda_{sc}^{++}) \mathcal{D}^-_\alpha - \left( \mathcal{D}^-_\alpha\Lambda_{sc}^{++} \right) \mathcal{D}^{--}
        \\  &-\left(\mathcal{D}^+_\alpha \Lambda_{sc}^{-\beta} \right) \mathcal{D}^-_\beta
        -
        \frac{1}{2}
        \left( \mathcal{D}^-_\beta  \Lambda_{sc}^{-\beta} \right) \mathcal{D}^+_\alpha\,,
    \end{split}
\end{equation}
\begin{equation}\label{eq: part sc}
    \begin{split}
        \delta_{sc} \partial_{\alpha\dot{\alpha}} = &
        -
        (\mathcal{D}^+_\alpha \Lambda_{sc}^{-\beta}) \partial_{\beta\dot{\alpha}}
        -
        (\bar{\mathcal{D}}^+_{\dot{\alpha}} \bar{\Lambda}_{sc}^{-\dot{\beta}}) \partial_{\alpha\dot{\beta}}
        -
        (\mathcal{D}^{--}\Lambda^{++}_{sc}) \partial_{\alpha\dot{\alpha}}
        \\
        &
        -
        \frac{i}{4} \left( \bar{\mathcal{D}}^-_{\dot{\alpha}} \Lambda_{sc}^{++} \right)
        \mathcal{D}^-_\alpha
        +
        \frac{i}{4} \left( \mathcal{D}^-_\alpha \Lambda_{sc}^{++} \right) \bar{\mathcal{D}}^-_{\dot{\alpha}}
        \\
        &
        - \frac{1}{2} \left( \partial_{\beta\dot{\alpha}} \Lambda_{sc}^{-\beta} \right)  \mathcal{D}^+_\alpha
        -
        \frac{1}{2}
        \left( \partial_{\alpha\dot{\beta}} \bar{\Lambda}_{sc}^{-\dot{\beta}} \right) \bar{\mathcal{D}}^+_{\dot{\alpha}}.
    \end{split}
\end{equation}
\end{subequations}

\subsection{Superconformal transformations of the spin-$\mathbf{3}$ superfields}

In order to deduce the superconformal transformation laws of the higher-spin prepotentials, we employ an indirect method based on the superconformal invariance
of the interacting hypermultiplet vertex \cite{Buchbinder:2024pjm}. Since the cubic vertex in the action \eqref{eq: hyper action + sc}
was constructed to respect the exact invariance
under $\mathcal{N}=2$ superconformal transformations, its total variation under $\delta_{sc}$
must vanish. By evaluating this variation with the help of the known transformation laws of the hypermultiplet fields,
we can determine the corresponding transformation rules of the higher-spin analytic operators and, thereby, of the components of $G^{++}$-superfields.

We start with a simple example that illustrates the main features of superconformal transformations of $\mathcal{N}=2$ supermultiplets described in Section \ref{sec: multiplets}.
It is the cubic vertex  involving the hypermultiplet in \eqref{eq: hyper action + sc}:
\begin{equation}\label{eq: spin 3 hyper}
S_{(3, \frac{1}{2}, \frac{1}{2})}
=
-\int d\zeta^{(-4)} q^{+a} \hat{\mathcal{H}}^{++}_{3} J q^+_a.
\end{equation}
By construction, this vertex is superconformally invariant. Being aware of
how superconformal symmetry acts on the hypermultiplet, one can restore the transformation of the spin-{\bf 3} operator.
The superconformal transformations of $q^+$ hypermultiplet read\footnote{Since we are working with an analytic superfield, we assume here that
the term with $\mathcal{D}^+_{\hat{\alpha}}$ derivatives in the operator \eqref{eq: Lambda operator} can be discarded.}
\begin{equation}
    \delta^*_{sc} q^{+}_a
    =
    - \frac{1}{2} \left\{ \lambda^M_{sc}, \partial_M \right\}_{\text{AGB}} q^+_a
    = - \hat{\Lambda}_{sc}   q^{+}_a - \frac{1}{2}\Omega_{sc} q^+_a,
\end{equation}
where
\begin{equation}
    \Omega_{sc} : = (-1)^M \partial_M \lambda_{sc}^{M}
    =
    \partial_{\alpha\dot{\alpha}}   \lambda_{sc}^{\alpha\dot{\alpha}}
    -
    \partial^-_\alpha    \lambda_{sc}^\alpha
    -
    \partial^-_{\dot{\alpha}}    \bar{\lambda}_{sc}^{\dot{\alpha}}
    +
    \partial^{--} \lambda^{++}.
\end{equation}
It will be preferable to work in the covariant basis, where
\begin{equation}
        \delta^*_{sc} q^{+}_a
    =
    - \frac{1}{2} \left\{ \Lambda^M_{sc}, \mathcal{D}_M \right\}_{\text{AGB}} q^+_a.
\end{equation}
Then the superconformal variation of the vertex \eqref{eq: spin 3 hyper} is given by
    \begin{equation}
    \delta_{sc} S_{(3, \frac{1}{2}, \frac{1}{2})}
        =
        -\int d\zeta^{(-4)} q^{+a}
        \left(
         \delta^*_{sc}\hat{\mathcal{H}}^{++}_{3}
         -
         \frac{1}{2}
        \left[\hat{\mathcal{H}}^{++}_{3} ,  \left\{ \Lambda^M_{sc}, \mathcal{D}_M \right\}_{\text{AGB}}  \right]
        \right)
        J q^+_a.
    \end{equation}
From this variation it is straightforward to determine the superconformal transformation laws of the prepotentials. One can check that superfields
with two multi-indices transform homogeneously, while the transformation law of the superfield without indices contains inhomogeneous terms.
The transformation law for the two-index superfields can be then determined to be:
\begin{equation}
 \delta^*_{sc} G^{++MN} \mathcal{D}_N \mathcal{D}_M
 =
    \left[\hat{\mathcal{H}}^{++}_{3} ,  \hat{\Lambda}_{sc} \right]   \Bigg|_{\text{two derivatives terms}}\,.
\end{equation}
Evaluating the commutator yields
\begin{equation}
    \delta^*_{sc} G^{++MN}
    =
    -\hat{\Lambda}_{sc} G^{++MN}
    +
    G^{++MK} \mathcal{D}_K \Lambda^N_{sc}
    +
    (-1)^{\epsilon_N [\epsilon_M+\epsilon_K]}
    G^{++KN} \mathcal{D}_K \Lambda^M _{sc}.
\end{equation}

From now on, it will be more convenient to deal with the passive variation, omitting the transport term.
To specify the superconformal transformations of the half-analytic superfield \eqref{eq: spin-3 HA} we consider\footnote{For compactness, in what follows we will delete the subscript 'sc', assuming that we are dealing with rigid $\mathcal{N}=2$ superconformal transformations.}
\begin{equation}
    \begin{split}
    \delta_{sc} G^{++\alpha\dot{\alpha} \beta\dot{\beta}}
    =&\,
    G^{++\alpha\dot{\alpha}\gamma\dot{\gamma} }
    \partial_{\gamma\dot{\gamma}} \Lambda^{\beta\dot{\beta}}
        +
    G^{++\gamma\dot{\gamma} \beta\dot{\beta}}
    \partial_{\gamma\dot{\gamma}}
 \Lambda^{\alpha\dot{\alpha}}
 \\
 &
 + G^{++\alpha\dot{\alpha} \gamma+} \mathcal{D}^-_{\gamma} \Lambda^{\beta\dot{\beta}}
  + G^{++ \gamma+ \beta\dot{\beta}} \mathcal{D}^-_{\gamma} \Lambda^{\alpha\dot{\alpha}}
   \\
  &
  +
  G^{++\alpha\dot{\alpha} \dot{\gamma}+} \bar{\mathcal{D}}^-_{\dot{\gamma}} \Lambda^{\beta\dot{\beta}}
  +
  G^{++ \dot{\gamma} + \beta\dot{\beta}} \bar{\mathcal{D}}^-_{\dot{\gamma}} \Lambda^{\alpha\dot{\alpha}}
  \\
  &
    +
  4i G^{++\alpha\dot{\alpha}\dot{\beta}+} \Lambda^{-\beta}
  +
  4i G^{++\beta\dot{\beta}\dot{\alpha}+} \Lambda^{-\alpha}
  \\
  &
    -
    4i  G^{++\alpha\dot{\alpha} \beta+}  \bar{\Lambda}^{-\dot{\beta}}
  -
  4i  G^{++ \alpha+ \beta\dot{\beta}}  \bar{\Lambda}^{-\dot{\alpha}}\,.
    \end{split}
\end{equation}
Taking into account the relations for the conformal superparameters collected in Section \ref{eq: sec rel Lambda}, we can rewrite this variation as
\begin{equation}
    \begin{split}
        \delta_{sc} G^{++\alpha\dot{\alpha} \beta\dot{\beta}}
        =&
        -
        2
        (\mathcal{D}^{--} \Lambda^{++})
        G^{++\alpha\dot{\alpha} \beta\dot{\beta} }
    %   \\
    %   &
        +
        G^{++\gamma\dot{\alpha}\beta \dot{\beta} }
        \mathcal{D}^-_{\gamma} \Lambda^{+\alpha}
        +
        G^{++\alpha\dot{\alpha}\gamma \dot{\beta} }
        \mathcal{D}^-_{\gamma} \Lambda^{+\beta}
    \\& +
        G^{++\alpha\dot{\gamma}\beta\dot{\beta}} \bar{\mathcal{D}}^-_{\dot{\gamma}} \Lambda^{+\dot{\alpha}}
            +
        G^{++\alpha\dot{\alpha}\beta\dot{\gamma}} \bar{\mathcal{D}}^-_{\dot{\gamma}} \Lambda^{+\dot{\beta}}.
    \end{split}
\end{equation}
Now we see that the symmetrized part transforms solely through the quantities with the symmetrized indices
\begin{equation}\label{eq: spin 3 vec sc}
\boxed{
        \delta_{sc} G^{++(\alpha \beta) (\dot{\alpha}\dot{\beta})}
        =\,
        2
        (\mathcal{D}^{--} \Lambda^{++})
    G^{++(\alpha \beta) (\dot{\alpha}\dot{\beta})}
        +
        2G^{++(\alpha\gamma)(\dot{\alpha} \dot{\beta}) }
        \mathcal{D}^+_{\gamma} \Lambda^{-\beta}
    +
        2G^{++(\alpha\beta)(\dot{\alpha}\dot{\gamma})} \bar{\mathcal{D}}^+_{\dot{\gamma}} \bar{\Lambda}^{-\dot{\beta}}.
 }
\end{equation}
Here we made use of the relations \eqref{eq: Lor factors}.

The completely analogous reasoning can be applied to the fermionic superfield.
Its transformation law takes the form
\begin{equation}
    \begin{split}
    \delta_{sc} G^{++\alpha\dot{\alpha}\beta+}
    =&\,
    \frac{i}{4} G^{++\alpha\dot{\alpha}\beta\dot{\gamma}} \bar{\mathcal{D}}^-_{\dot{\gamma}} \Lambda^{++}
    +
    2(\mathcal{D}^{--} \Lambda^{++}) G^{++\alpha\dot{\alpha}\beta+}
    \\
&   +
    G^{++\alpha\dot{\gamma}\beta+} \bar{\mathcal{D}}^+_{\dot{\gamma}} \Lambda^{-\dot{\beta}}
    +
    G^{++\gamma\dot{\beta} \beta+} \mathcal{D}^+_\gamma \Lambda^{-\alpha}
    +
     G^{++\alpha\dot{\beta}\gamma+} \mathcal{D}^+_\gamma \Lambda^{-\beta}\,,
        \end{split}
\end{equation}
and, after symmetrization and making use of the relations \eqref{eq: Lor factors}, becomes
\begin{equation}\label{eq: spin 3 spin sc}
\boxed{     \begin{split}
            \delta_{sc} G^{++(\alpha\beta)\dot{\alpha}+}
        =&\,
        \frac{i}{4} G^{++(\alpha\beta)(\dot{\alpha}\dot{\gamma})} \bar{\mathcal{D}}^-_{\dot{\gamma}} \Lambda^{++}
        +
        2(\mathcal{D}^{--} \Lambda^{++}) G^{++(\alpha\beta)\dot{\alpha}+}
        \\
        &   +
        G^{++(\alpha\beta)\dot{\gamma}+}  \bar{\mathcal{D}}^+_{\dot{\gamma}} \Lambda^{-\dot{\alpha}}
        +
        2
        G^{++(\gamma\beta)\dot{\beta} +} \mathcal{D}^+_\gamma \Lambda^{-{\alpha}}.
    \end{split} }
\end{equation}

The  transformation laws obtained in this way are a straightforward generalization of the superconformal transformation laws for the spin $\mathbf{2}$ case
presented in ref. \cite{Ivanov:2025gvs}.

\subsection{Superconformal invariance of $\mathcal{N}=2$ spin-$\mathbf{3}$ Weyl action}

To check superconformal invariance of the action \eqref{eq: spin s action}, we will use the method applied in ref. \cite{Ivanov:2025gvs}. We will work out in detail
the simplest spin-$\mathbf{3}$ example, using the transformation laws \eqref{eq: spin 3 vec sc} and \eqref{eq: spin 3 spin sc}.
The superconformal transformation of the half-analytic superfield $\mathcal{H}^{++}_{\alpha(4)}$
is driven by the superconformal parameters $\Lambda^{-\beta}$, their conjugate partners $\bar{\Lambda}^{-\dot{\alpha}}$, and the parameter $\Lambda^{++}$.
To check whether the action is invariant, we analyze the superconformal variations with these parameters sector by sector.
Then, finally, we assemble the whole variation from these constituents.

\smallskip

\noindent \underline{$\Lambda^{-\beta}$ sector}

\smallskip

The variation with the parameter $\Lambda^{-\beta}$ yields:
\begin{equation}
    \delta^{(1)}_{sc} \mathcal{H}^{++}_{\alpha(4)}
    =
    - 4 \left(\mathcal{D}^+_{(\alpha} \Lambda^{-\beta)} \right)
    \mathcal{H}^{++}_{(\beta\alpha(3))}
    -
    \left(
     \partial_\alpha^{\dot{\beta}} \mathcal{D}^+_\alpha \Lambda^{-\beta} \right)
    \left(
    \partial_\beta^{\dot{\beta}} G^{++}_{\alpha(2)\dot{\beta}(2)}
    +
    2
    \mathcal{D}^-_\beta G^{+++}_{\alpha(2)\dot{\beta}}
    \right).
\end{equation}
The first term makes a homogeneous contribution to the transformation law, while it is more convenient to preliminary bring the second term in another form.
We use the identity for the superconformal parameters,
\begin{equation}
     \partial_\alpha^{\dot{\beta}} \mathcal{D}^+_\alpha \lambda^{-\beta}
     =
     \frac{i}{4} \delta_\alpha^\beta \mathcal{D}^-_\alpha \bar{\mathcal{D}}^{-\dot{\beta}} \Lambda^{++},
\end{equation}
and find the final form of the variation as
\begin{equation}
    \delta^{(1)}_{sc} \mathcal{H}^{++}_{\alpha(4)}
    =
    - 4 \left(\mathcal{D}^-_{(\alpha} \Lambda^{+\beta)} \right)
    \mathcal{H}^{++}_{(\beta\alpha(3))}
    -
    \frac{i}{4}
    \left(
\mathcal{D}^-_\alpha \bar{\mathcal{D}}^{-\dot{\beta}} \Lambda^{++}
 \right)
    \left(
    \partial_\alpha^{\dot{\beta}} G^{++}_{\alpha(2)\dot{\beta}(2)}
    +
    2
    \mathcal{D}^-_\alpha G^{+++}_{\alpha(2)\dot{\beta}}
    \right).
\end{equation}

\noindent \underline{$\bar{\Lambda}^{-\dot{\beta}}$ sector}

\smallskip

Similarly, computing the variation in the $\bar{\Lambda}^{-\dot{\beta}}$ sector and using the relation
\begin{equation}
    \partial_\alpha^{\dot{\alpha}} \bar{\mathcal{D}}^{+\dot{\beta}} \bar{\Lambda}^-_{\dot{\rho}}
    =
    \frac{1}{2} \delta^{\dot{\alpha}}_{\dot{\rho}}
    \bar{\mathcal{D}}^{+\dot{\beta}} \left( \partial_\alpha^{\dot{\sigma}} \bar{\Lambda}_{\dot{\sigma}}^- \right)
    =
    -
    \frac{i}{4} \left( \mathcal{D}^-_\alpha \bar{\mathcal{D}}^{-\dot{\beta}} \Lambda^{++} \right),
\end{equation}
we obtain
\begin{equation}
    \begin{split}
    \delta^{(2)}_{sc} \mathcal{H}^{++}_{\alpha(4)}
    =&
    + \frac{5i}{4} \left(  \mathcal{D}^-_\alpha \bar{\mathcal{D}}^{-\dot{\beta}}  \Lambda^{++} \right)  \partial_\alpha^{\dot{\beta}} G^{++}_{\alpha(2) \dot{\beta}(2)}
\\
&
%
%-16i  (\partial_{\alpha\dot{\rho}} \bar{\Lambda}^{-\dot{\rho}}) \partial_\alpha^{\dot{\beta}} G^{+++}_{\alpha(2)\dot{\beta}}
+
i \left( \mathcal{D}^-_\alpha \bar{\mathcal{D}}^{-\dot{\beta}} \Lambda^{++} \right)  \mathcal{D}^-_\alpha G^{+++}_{\alpha(2)\dot{\rho}}
-
6 \left( \mathcal{D}^-_\alpha \mathcal{D}^{--} \Lambda^{++} \right)
\partial_\alpha^{\dot{\beta}} G^{+++}_{\alpha(2)\dot{\beta}}.
\end{split}
\end{equation}

\noindent \underline{$\Lambda^{++}$ sector}

\smallskip

In this case, a careful calculation yields
\begin{equation}
    \begin{split}
    \delta_{sc}^{(3)}
    \mathcal{H}^{++}_{\alpha(4)}
    =&\;
    \frac{i}{2} \left( \mathcal{D}^-_\alpha \Lambda^{++} \right) \bar{\mathcal{D}}^{-\dot{\beta}} \partial^{\dot{\beta}}_{\alpha} G^{++}_{\alpha(2)\dot{\beta}(2)}
    -
    \frac{i}{2} \left( \mathcal{D}^-_\alpha \bar{\mathcal{D}}^{-\dot{\beta}} \Lambda^{++} \right)
    \partial_\alpha^{\dot{\beta}} G^{++}_{\alpha(2)\dot{\beta}(2)}\
    \\
    &
    +
    \frac{i}{2} \left( \mathcal{D}^-_\alpha \Lambda^{++}  \right) \bar{\mathcal{D}}^{-\dot{\beta}} \mathcal{D}^-_\alpha G^{+++}_{\alpha(2)\dot{\beta}}
    \\
    &
    +
    4\left(\mathcal{D}^-_\alpha \mathcal{D}^{--} \Lambda^{++} \right)
    \partial_\alpha^{\dot{\beta}} G^{+++}_{\alpha(2)\dot{\beta}}
    -
    2 \left( \mathcal{D}^-_\alpha \Lambda^{++} \right)
    \mathcal{D}^{--}
    \partial_\alpha^{\dot{\beta}}  G^{+++}_{\alpha(2)\dot{\beta}}.
    \end{split}
\end{equation}

\noindent \underline{Net result}

Summing up the above contributions, we find the total variation in the form
\begin{equation}\label{VARbas}
    \begin{split}
    \delta_{sc} \mathcal{H}^{++}_{\alpha(4)}
    =
    &
        - 4 \left(\mathcal{D}^+_{(\alpha} \Lambda^{-\beta)} \right)
    \mathcal{H}^{++}_{(\beta\alpha(3))}
\\& -
    \frac{i}{2}
    \bar{\mathcal{D}}^{-\dot{\beta}}
    \left[
    \left( \mathcal{D}^-_\alpha \Lambda^{++} \right)
    \left(\partial^{\dot{\beta}}_{\alpha} G^{++}_{\alpha(2)\dot{\beta}(2)}
    +
    \mathcal{D}^-_\alpha G^{+++}_{\alpha(2)\dot{\beta}}
     \right)
    \right]
\\
&       -
    2
        \mathcal{D}^{--}
        \left[
     \left( \mathcal{D}^-_\alpha \Lambda^{++} \right)
    \partial_\alpha^{\dot{\beta}}  G^{+++}_{\alpha(2)\dot{\beta}}
    \right].
    \end{split}
\end{equation}

Consider the inhomogeneous terms. To this end, we introduce the following notation:
\begin{equation}
    \begin{split}
&X^{(+4)}_{\alpha(4)} = -2  \left[
    \left( \mathcal{D}^-_\alpha \Lambda^{++} \right)
    \partial_\alpha^{\dot{\beta}}  G^{+++}_{\alpha(2)\dot{\beta}}
    \right],
\\
&   Y^{(+3)}_{\alpha(4)\dot{\beta}}
    =
    \frac{i}{2}
      \left[
    \left( \mathcal{D}^-_\alpha \Lambda^{++} \right)
    \left(\partial^{\dot{\beta}}_{\alpha} G^{++}_{\alpha(2)\dot{\beta}(2)}
    +
    \mathcal{D}^-_\alpha G^{+++}_{\alpha(2)\dot{\beta}}
    \right)
    \right],
    \\
&
K^{(+4)}_{\alpha(4)}
=
-   \frac{i}{2} \bar{\mathcal{D}}^{+\dot{\beta}}
\left[
\left( \mathcal{D}^-_\alpha \Lambda^{++} \right)
\left(\partial^{\dot{\beta}}_{\alpha} G^{++}_{\alpha(2)\dot{\beta}(2)}
+
\mathcal{D}^-_\alpha G^{+++}_{\alpha(2)\dot{\beta}}
\right)
\right].
    \end{split}
\end{equation}
Then, using the harmonic zero-curvature equations,
\begin{equation}
    \mathcal{D}^{--} X^{(+4)}_{\alpha(4)} =     \mathcal{D}^{++}  X^{(0)}_{\alpha(4)},
    \qquad
        \mathcal{D}^{--}  Y^{(+3)}_{\alpha(4)\dot{\beta}}
        =
            \mathcal{D}^{++}  Y^{-}_{\alpha(4)\dot{\beta}},
            \qquad
        \mathcal{D}^{--}    K^{(+4)}_{\alpha(4)}
        =
            \mathcal{D}^{++}    K^{(0)}_{\alpha(4)}\,, \label{Zero-modes}
\end{equation}
     we can represent the inhomogeneous contributions to \eqref{VARbas} as
\begin{equation}\label{eq: sc inh}
    \delta_{sc} \mathcal{H}^{++}_{\alpha(4)} \Big|_{\text{inhom}}
    =
    \mathcal{D}^{++} \left(
     X^{(0)}_{\alpha(4)}
    +
     \bar{\mathcal{D}}^{+\dot{\beta}} Y^{-}_{\alpha(4)\dot{\beta}} +
     K^{(0)}_{\alpha(4)}     \right).
\end{equation}
If the expression inside the bracket in \eqref{eq: sc inh} were  half-analytic, this variation would be a particular form of the gauge transformation \eqref{eq: spin 3 gauge}.
The sought half-analyticity can be achieved by noting that the gauge parameter in \eqref{eq: sc inh} is defined modulo a term $Z^{(0)}_{\alpha(4)}$,
such that $$\mathcal{D}^{++}Z^{(0)}_{\alpha(4)} = 0.$$
This freedom amounts to the presence of zero-modes in the solution of the harmonic zero-curvature conditions \eqref{Zero-modes} (of the first and third ones).
Though this addition of the parameter $Z^{(0)}_{\alpha(4)}$ leaves the transformation law  \eqref{eq: sc inh} unaffected,
the modified gauge parameter can be chosen
half-analytic,
\begin{equation}\label{halfAnalLambda}
    \Lambda^{(sc)}_{\alpha(4)} = Z^{(0)}_{\alpha(4)}
    +
         X^{(0)}_{\alpha(4)}
    +
    \bar{\mathcal{D}}^{+\dot{\beta}} Y^{-}_{\alpha(4)\dot{\beta}} +
    K^{(0)}_{\alpha(4)},
    \qquad
    \bar{\mathcal{D}}^{+}_{\dot{\rho}}  \Lambda^{(sc)}_{\alpha(4)}  = 0\,.
\end{equation}
Details are given in the appendix \ref{App: half-analytic}.
Thereby we have demonstrated that the inhomogeneous terms are reduced to gauge transformations.

\medskip

The conclusion is that, modulo the gauge transformation $\mathcal{D}^{++}\Lambda_{\alpha(4)}^{(sc)} $, the final rigid superconformal transformation law ca be written as:
\begin{equation}
    \boxed{ \delta_{sc} \mathcal{H}^{++}_{\alpha(4)}
    =
    - 4 \left(\mathcal{D}^+_{(\alpha} \Lambda_{sc}^{-\beta)} \right)
    \mathcal{H}^{++}_{(\beta\alpha(3))}. }
\end{equation}
Taking into account the superconformal transformation law of the harmonic derivatives \eqref{eq: N=2 sc 1} and \eqref{eq: N=2 sc 2}, we can also deduce the transformation law
of the negatively charged superfield,
\begin{equation}
    \boxed{ \delta_{sc} \mathcal{H}^{--}_{\alpha(4)}
        =
        - 4 \left(\mathcal{D}^+_{(\alpha} \Lambda_{sc}^{-\beta)} \right)
        \mathcal{H}^{++}_{(\beta\alpha(3))}
        -
        \left( \mathcal{D}^{--} \Lambda^{++}_{sc} \right) \mathcal{H}^{--}_{\alpha(4)} . }
\end{equation}
The first term looks as a homogeneous rotation with respect to each of the spinor indices; the second term
 is a weight transformation.

Having established this transformation law, we are ready to evaluate the variation of the spin-$\mathbf{3}$ super Weyl tensor.
To this end, we must account for the transformation of the chiral projecting operator itself, whose superconformal variation generates
specific weight contributions, as dictated by $\delta_{sc} \bar{\mathcal{D}}^+_{\dot{\alpha}} =
- \left( \bar{\mathcal{D}}^+_{\dot{\alpha}} \Lambda^{-\dot{\beta}} \right) \bar{\mathcal{D}}^+_{\dot{\beta}}$.
Combining these contributions with $\delta_{sc} \mathcal{H}^{--}_{\alpha(4)}$, we arrive at the variation:
\begin{equation}
    \boxed{ \delta_{sc} \mathcal{W}_{\alpha(4)}
    =
        - 4 \left(\mathcal{D}^+_{(\alpha} \Lambda^{-\beta)}_{sc} \right)
         \mathcal{W}_{(\alpha(3)\beta)}
         -
         \left( \bar{\mathcal{D}}^+_{\dot{\beta}} \bar{\Lambda}_{sc}^{-\dot{\beta}} \right)   \mathcal{W}_{\alpha(4)}
         -
         \left( \mathcal{D}^{--} \Lambda^{++}_{sc} \right)
         \mathcal{W}_{\alpha(4)}. }
\end{equation}
The invariance of the spin-$\mathbf{3}$ action \eqref{eq: spin 3 action} straightforwardly follows  from both this formula and the superconformal transformation
rules of the full HSS and chiral superspace measures:
\begin{equation}\label{eq: sc measures}
    \begin{split}
        & \delta_{sc} \left( d^4x d^8\theta du \right) =
             \left[ \mathcal{D}^{--} \Lambda^{++}_{sc} \right]
        \left( d^4 xd^8\theta du \right),
        \\
        &
        \delta_{sc} \left( d^4x_L d^4 \theta  \right)
        =
        2\left[ \bar{\mathcal{D}}^+_{\dot{\alpha}} \bar{\Lambda}_{sc}^{-\dot{\alpha}} +  \mathcal{D}^{--} \Lambda^{++}_{sc}  \right]
        \left( d^4x_L d^4 \theta  \right).
    \end{split}
\end{equation}
Indeed, the superconformal  transformations of the corresponding Lagrangian densities,
\begin{subequations}
\begin{equation}
    \delta_{sc} \left(
    \mathcal{H}^{++\alpha(4)} \mathcal{H}^{--}_{\alpha(4)}
     \right)
     =
        -
     \left( \mathcal{D}^{--} \Lambda^{++}_{sc} \right)
     \left(
     \mathcal{H}^{++\alpha(4)} \mathcal{H}^{--}_{\alpha(4)}
     \right),
\end{equation}
\begin{equation}
    \delta_{sc}
    \left( \mathcal{W}^{\alpha(4)}  \mathcal{W}_{\alpha(4)} \right)
    =
-   2\left( \bar{\mathcal{D}}^+_{\dot{\alpha}} \bar{\Lambda}_{sc}^{-\dot{\alpha}} +  \mathcal{D}^{--} \Lambda^{++}_{sc}  \right)
        \left( \mathcal{W}^{\alpha(4)}  \mathcal{W}_{\alpha(4)} \right),
\end{equation}
\end{subequations}
exactly cancel \eqref{eq: sc measures}.

Now we turn to the similar analysis of the general spin-$\mathbf{s}$ case.

\subsection{Superconformal invariance of the integer spin-$\mathbf{s}$ Weyl action}\label{sec: gen spin s sc}

An  effect of intricate cancelation of the inhomogeneous terms observed in the spin-$\mathbf{3}$ case naturally extends
to an arbitrary integer spin $\mathbf{s}$. While the intermediate calculations become increasingly tedious due to the presence of $(s-1)$ space-time
derivatives in the definition of the higher-spin half-analytic superfield $\mathcal{H}^{++}_{\alpha(2s-2)}$, the underlying logic scheme remains the same.

By performing the superconformal variations and discarding the terms that reduce to gauge transformations $\mathcal{D}^{++}\Lambda_{\alpha(2s-2)}^{(sc)}$,
one finds that the final transformation law for the half-analytic prepotential is also homogeneous. It generalizes the $s=3$ result to the total number $2s-2$ of undotted indices:
\begin{subequations}
\begin{equation}\label{eq: spin s prepotential sc}
    \delta_{sc} \mathcal{H}^{++}_{\alpha(2s-2)} = - (2s-2) \left(\mathcal{D}^+_{(\alpha} \Lambda^{-\beta)}_{sc} \right) \mathcal{H}^{++}_{(\beta\alpha(2s-3))},
\end{equation}
\begin{equation}\label{eq: spin s-- prepotential sc}
    \delta_{sc} \mathcal{H}^{--}_{\alpha(2s-2)} = - (2s-2) \left(\mathcal{D}^+_{(\alpha} \Lambda^{-\beta)}_{sc} \right) \mathcal{H}^{--}_{(\beta\alpha(2s-3))}
    -
    (\mathcal{D}^{--}\Lambda^{++}_{sc} )  \mathcal{H}^{--}_{\alpha(2s-2)}.
\end{equation}
\end{subequations}
The transformation laws involve a homogeneous rotation with respect to each of the $(2s -2)$ symmetrized spinor indices and the weight terms.

The corresponding Lagrangian density has the universal transformation law
\begin{equation}
    \delta_{sc} \left(  \mathcal{H}^{++\alpha(2s-2)}  \mathcal{H}^{--}_{\alpha(2s-2)} \right)
    =
        -
    \left(\mathcal{D}^{--}\Lambda^{++}_{sc} \right)
    \left(  \mathcal{H}^{++\alpha(2s-2)}  \mathcal{H}^{--}_{\alpha(2s-2)} \right) .
\end{equation}
Due to the transformation law \eqref{eq: sc measures} of the full HSS integration measure, the action \eqref{eq: spin s action} is superconformally invariant.

\medskip

To determine the superconformal variation of the corresponding Weyl tensor $\mathcal{W}_{\alpha(2s-2)} = (\bar{\mathcal{D}}^+)^2 \mathcal{H}^{--}_{\alpha(2s-2)}$,
one needs to evaluate the action of the superconformal group on the chiral projector. As demonstrated in the previous subsection, this procedure generates
additional scaling weight contributions. For an arbitrary integer spin $s$, the complete transformation law takes the universal form:
\begin{equation}\label{eq: spin s Weyl tensor sc}
    \begin{split}
        \delta_{sc} \mathcal{W}_{\alpha(2s-2)} =\;& - (2s-2) \left(\mathcal{D}^+_{(\alpha} \Lambda^{-\beta)}_{sc} \right) \mathcal{W}_{(\beta \alpha(2s-3))} \\
        & - \left( \bar{\mathcal{D}}^+_{\dot{\beta}} \bar{\Lambda}_{sc}^{-\dot{\beta}} \right) \mathcal{W}_{\alpha(2s-2)} - \left( \mathcal{D}^{--} \Lambda^{++}_{sc} \right) \mathcal{W}_{\alpha(2s-2)}.
    \end{split}
\end{equation}
The notable feature of this result is that terms with the scalar weight factors, {\it i.e.}  the last two terms in \eqref{eq: spin s Weyl tensor sc}, are
same for any value of the spin $s$. This universality is by no means a coincidence, it deeply roots in the chiral superspace structure of the final action.

The higher-spin Weyl action \eqref{eq: spin s action} can be alternatively written as an integral over the $\mathcal{N}=2$ chiral superspace,
\begin{equation*}
    S_{\text{Weyl}}^s = \int d^4x_L d^4\theta \ \mathcal{W}^{\alpha(2s-2)} \mathcal{W}_{\alpha(2s-2)} + \text{c.c.}\,.
\end{equation*}
For the invariance of this action, the Lagrangian density $\mathcal{L}_c = \mathcal{W}^{\alpha(2s-2)} \mathcal{W}_{\alpha(2s-2)}$ must transform
as a proper chiral density exactly canceling the superconformal variation of the chiral integration measure $d^4x_L d^4\theta$.

To see that it is indeed the case, we note that, when computing the variation of the quadratic contraction, the terms with mixed indices are perfectly canceled out among themselves.
This happens due to the property that the operator $\mathcal{D}^+_{(\alpha} \Lambda^{-\beta)}_{sc}$ is symmetric in its indices and so effectively acts
as a traceless $\mathfrak{sl}(2,\mathbb{C})$ rotation preserving the pair of contracted indices. Thus the full contraction
$\mathcal{W}^{\alpha(2s-2)} \mathcal{W}_{\alpha(2s-2)}$ behaves as a Lorentz scalar with respect to these rotations.
As a result, the variation of the chiral Lagrangian density is reduced to the universal weight transformation,
%which simply double upon varying the quadratic product:
\begin{equation}
    \delta_{sc} \left( \mathcal{W}^{\alpha(2s-2)} \mathcal{W}_{\alpha(2s-2)} \right) = - 2 \left(\bar{\mathcal{D}}^+_{\dot{\beta}} \bar{\Lambda}_{sc}^{-\dot{\beta}}  +
    \mathcal{D}^{--} \Lambda^{++}_{sc} \right) \mathcal{W}^{\alpha(2s-2)} \mathcal{W}_{\alpha(2s-2)}.
\end{equation}
This overall scalar factor is precisely what one needs  for canceling the superconformal variation of the chiral measure \eqref{eq: sc measures}.
This proves $\mathcal{N}=2$ superconformal invariance of the action $S_{\text{Weyl}}^s$ for any integer spin $s$.

The general reasoning we have used here circumvents the necessity of explicit calculations and completes the proof of the rigid superconformal invariance
of the entire hierarchy of $\mathcal{N}=2$ higher-spin generalized Weyl theories.

\section{$\mathcal{N} =2$ supergravity background}\label{sec: sugra back}

We now turn to inspecting the possibility to formulate the theory on $\mathcal{N} =2$ supergravity background.

We begin by recalling \cite{Buchbinder:2024pjm} that the hypermultiplet interaction with the spin $\mathbf{2}$
and spin $\mathbf{s}$ supermultiplets constructed as
\begin{equation}\label{eq: 2+s}
    S = -\frac{1}{2} \int d\zeta^{(-4)} q^{+a} \left( \mathcal{D}^{++} + \kappa_2 \hat{\mathcal{H}}^{++}_{s=2} + \kappa_s \hat{\mathcal{H}}^{++}_s J^{P(s)} \right) q^+_a,
\end{equation}
is invariant under arbitrary conformal supergravity transformations after a suitable non-Abelian modification of the free transformation law  \eqref{eq: transf conv} for the spin $\mathbf{2}$
and spin $\mathbf{s}$ analytic operators. This follows from the property that the commutator of a differential super-operator
of degree $1$ (which corresponds to local superconformal transformations) with an operator of degree $s$, again yields  a differential operator of the  degree  $s$.
The corresponding local transformation law of the spin-$\mathbf{s}$ operator takes the form
\begin{equation}
    \delta_{lsc} \hat{\mathcal{H}}^{++}_s
    =
    [\hat{\mathcal{H}}^{++}_s, \hat{\mathcal{U}}_2].
\end{equation}
So the spin-$\mathbf{s}$ gauge transformation gets non-Abelian corrections,
    \begin{equation}\label{eq: spin s NA mod}
    \delta^{(s,k)}_\lambda  \hat{\mathcal{H}}^{++}_{s} =
    \frac{1}{2} [\mathcal{D}^{++}, \hat{\mathcal{U}}_k ]
    \quad
    \to
    \quad
    \delta^{(s,k)}_\lambda  \hat{\mathcal{H}}^{++}_{s} =
    \frac{1}{2} [\mathcal{D}^{++} + \kappa_2 \hat{\mathcal{H}}^{++}_{2}, \hat{\mathcal{U}}_k ],
\end{equation}
which mix the spin-$\mathbf{s}$ and spin-$\mathbf{2}$ superfields.

The above observation has an important consequence. Since the hypermultiplet
action \eqref{eq: 2+s} is invariant under the local superconformal
transformations with the non-Abelian modification \eqref{eq: spin s NA mod},
the logarithmically divergent part of the induced effective action
\begin{equation}
    e^{iS_{ind}} = \int \mathfrak{D} q^+ \mathfrak{D} \tilde{q}^+
    e^{-\frac{i}{2} \int d\zeta^{(-4)} q^{+a} \left( \mathcal{D}^{++} + \kappa_2 \hat{\mathcal{H}}^{++}_{s=2} + \kappa_s \hat{\mathcal{H}}^{++}_s J^{P(s)} \right) q^+_a}
\end{equation}
 is invariant under the local superconformal transformations of
the full nonlinear $\mathcal{N}=2$ supergravity. This shows that
\eqref{eq: spin s NA mod} is a consistent gauge transformation law for the
effective action quadratic in the higher-spin superfields on an arbitrary
$\mathcal{N}=2$ supergravity background. One might then try to construct a
classical action with the same property by a straightforward covariantization
of the flat Weyl action \eqref{eq: spin s action}, with the flat gauge
transformation law \eqref{eq: H++ gauge} covariantized in the direct way.

A closer look shows, however, that this expectation is too naive. A similar
phenomenon was observed in ref.~\cite{Beccaria:2017nco} for the
conformal spin-$3$ field on a curved gravitational background: the spin-$3$
gauge invariance required additional spin-$1$/spin-$3$ mixing terms in
the gauge transformation law, {\it i.e.} a deformation of the naive covariantization,
rather than directly using the latter. Motivated by this analogy, one may expect
that the complete gauge-invariant coupling of $\mathcal{N}=2$ higher-spin
supermultiplets to supergravity will also require some additional mixing terms
and/or a deformation of the gauge transformation law.

In this section, following our previous work \cite{Ivanov:2025gvs}, we
perform the covariantization of the spin-$s$ Weyl action \eqref{eq: spin s
    action} on  $\mathcal{N}=2$ supergravity background, assuming {\it a priori} no any restrictions on the latter. We show
that, while the resulting action is invariant under local superconformal
transformations, its spin-$s$ gauge invariance is restricted to $\mathcal{N}=2$
Weyl-flat superbackgrounds.
This conclusion is consistent with the results obtained in the framework
of the conformal superspace approach \cite{Kuzenko:2021pqm}.
This confirms the above expectation and indicates that constructing the gauge-invariant coupling on a generic supergravity background
would entail the proper  deformation of the gauge transformation law. The non-Abelian
transformation law \eqref{eq: spin s NA mod} suggests a possible form of
such a deformation; we leave detailed analysis of this opportunity for the future work.

\subsection{$\mathcal{N}=2$ Weyl supermultiplet and hybrid basis in curved HSS}\label{sec: hybr basis}

We start from the HSS description of the \(\mathcal{N}=2\) Weyl supermultiplet \cite{18, Galperin:1987ek, Ivanov:2022vwc}.
In the harmonic framework this multiplet is carried by the first-order analytic super-operator \(\hat{\mathcal{H}}^{++}_{(s=2)}\),
which properly covariantizes the harmonic derivative\footnote{A detailed discussion of WZ gauge for \(\mathcal{N}=2\) Weyl supermultiplet
at the nonlinear level can be found in ref. \cite{Zaigraev:2026gjb}.}
\begin{equation}\label{eq: Weyl prep}
    \begin{split}
    \mathfrak{D}^{++} =& \, \mathcal{D}^{++} + \kappa_2 \hat{\mathcal{H}}^{++}_{(s=2)}
    \\
    =&\,  \partial^{++}
    +
    \mathcal{H}^{++\alpha\dot{\alpha}} \partial_{\alpha\dot{\alpha}}
    +
    \mathcal{H}^{++\alpha+} \partial^-_\alpha
        +
    \mathcal{H}^{++\dot{\alpha}+} \partial^-_{\dot{\alpha}}
    +
    \mathcal{H}^{(+4)} \partial^{--}.
    \end{split}
\end{equation}
The gauge group of $\mathcal{N}=2$ Weyl supergravity is generated by arbitrary analyticity-preserving coordinate transformations
\begin{equation}
\begin{split}
    &\delta x^{\alpha\dot{\alpha}}
    =
    \lambda^{ \alpha\dot{\alpha}}(\zeta).
    \\
    &
    \delta \theta^{+\hat{\alpha}}
    =
    \lambda^{ +\hat{\alpha}} (\zeta),
    \\
    &
        \delta \theta^{ -\hat{\alpha}}
    =
    \lambda^{-\hat{\alpha}}
    (\zeta, \theta^-),
    \\
    &
   \delta  u^{+}_i = \lambda^{++}(\zeta) u^-_i,
    \\
    &
     \delta u^{-}_i =  0.
\end{split}
\end{equation}
These transformations are the analytic localization of the rigid superconformal transformations from the section \ref{eq: sec rel Lambda};
we will refer to them as \textit{local superconformal transformations}.

With respect to these transformations, the harmonic derivative transforms according to
\begin{equation}
    \delta_\lambda \mathfrak{D}^{++}
    =
    -\lambda^{++} \mathcal{D}^0, \label{TranConfD++}
\end{equation}
which amounts to the following transformation laws for the prepotentials
\begin{equation}\label{eq: H transf}
    \begin{split}
        \delta_\lambda \mathcal{H}^{++m} &= \mathfrak{D}^{++}  \lambda^{m},
        \\
        \delta_\lambda \mathcal{H}^{++\hat{\alpha}+} &= \mathfrak{D}^{++} \lambda^{+\hat{\alpha}}
        -
        \lambda^{++} \theta^{+\hat{\alpha}},
        \\
        \delta_\lambda \mathcal{H}^{++\hat{\alpha}-} &= \mathfrak{D}^{++} \lambda^{-\hat{\alpha}} + \lambda^{++} \theta^{-\hat{\alpha}},
        \\
        \delta_\lambda
        \mathcal{H}^{(+4)} &= \mathfrak{D}^{++} \lambda^{++}.
    \end{split}
\end{equation}
The harmonic derivative of the opposite charge,
\begin{equation}
    \mathfrak{D}^{--} = \partial^{--}
    +
        \mathcal{H}^{--\alpha\dot{\alpha}} \partial_{\alpha\dot{\alpha}}
    +
    \mathcal{H}^{--\alpha+} \partial^-_\alpha
    +
    \mathcal{H}^{--\dot{\alpha}+} \partial^-_{\dot{\alpha}}
        +
    \mathcal{H}^{--\alpha-} \partial^+_\alpha
    +
    \mathcal{H}^{--\dot{\alpha}-} \partial^+_{\dot{\alpha}},
\end{equation}
is specified by the nonlinear zero-curvature condition:
\begin{equation}\label{eq: SG full harm}
    [\mathfrak{D}^{++} - \mathcal{H}^{(+4)}\mathfrak{D}^{--} ,  \mathfrak{D}^{--}]
    =
    \mathcal{D}^0.
\end{equation}
The condition \eqref{eq: SG full harm} is covariant  under the transformations \eqref{TranConfD++}, \eqref{eq: H transf} and
\begin{equation}\label{eq: mathfrak D trans}
\delta_\lambda \mathfrak{D}^{--} = - \left( \mathfrak{D}^{--} \lambda^{++}\right) \mathfrak{D}^{--}.
\end{equation}
   % \begin{split}
   %     &\delta_\lambda \mathfrak{D}^{++}
   %     =
   %     -\lambda^{++} \mathcal{D}^0,
   %     \\
    %    &\delta_\lambda \mathfrak{D}^{--}
    %    =
    %    -
    %    \left( \mathfrak{D}^{--} \lambda^{++}\right) \mathfrak{D}^{--},
    %    \\
    %    &   \delta_\lambda \mathcal{H}^{(+4)}
    %    =
    %    \mathfrak{D}^{++} \lambda^{++}.
    %\end{split}

In order to construct invariant actions, we should also introduce the density $E$ ensuring the homogeneous transformation law for the integration measure:
\begin{equation}
    \delta_\lambda \left(d^4x d^8\theta du \, E\right)
    =
    \left( \mathfrak{D}^{--} \lambda^{++} \right) \left(d^4x d^8\theta du \, E\right).
\end{equation}
The density $E$ can be expressed in terms of the fundamental analytic $\mathcal{N}=2$ supergravity prepotentials defined above
(see refs. \cite{Galperin:1987ek, Galperin:1987em, 18, Ivanov:2022vwc} for details).

\medskip

One of the basic ingredients of the HSS construction of superconformal higher-spin theories is the homogeneous transformation law \eqref{eq: D+ sc}
at least for one of the short spinor derivatives, e.g.,
\begin{equation}
    \delta_{sc} \bar{\mathcal{D}}^+_{\dot{\alpha}}
    =
    - \left(\bar{\mathcal{D}}^+_{\dot{\alpha}} \bar{\lambda}^{-\dot{\beta}}_{sc} \right) \bar{\mathcal{D}}^+_{\dot{\beta}}.
\end{equation}
This law follows from the rigid \(\mathcal{N}=2\) superconformal relation \(\bar{\mathcal{D}}^+_{\dot{\alpha}} \lambda^{-\beta}_{sc} = 0\) and guarantees
that the concept of half-analytic superfields respects rigid superconformal symmetry. However, for the generic analyticity-preserving coordinate transformations,
which correspond to general local superconformal transformations, this property fails to hold,
\begin{equation}
    \delta_{lsc} \bar{\partial} ^+_{\dot{\alpha}}
    =
    -
    \left( \bar{\partial}^+_{\dot{\alpha}} \lambda^{-\beta}\right)
    \partial^+_{\beta}
    -
    \left(\bar{\partial}^+_{\dot{\alpha}} \bar{\lambda}^{-\dot{\beta}} \right) \bar{\partial}^+_{\dot{\beta}}.
\end{equation}

However, as observed in \cite{Galperin:1987em}, one can introduce a \textit{hybrid basis} in the curved harmonic superspace,
\begin{equation} \label{Hybrid}
    Z_H = \{ x_H^{\alpha\dot{\alpha}}, \theta_H^{\pm\hat{\alpha}}, u^\pm_i \}
    =
    \{ x_A^{\alpha\dot{\alpha}}, \theta_A^{+\hat{\alpha}}, \bar{\theta}_A^{-\dot{\alpha}},  \theta_H^{-\alpha} := \theta_A^{-\alpha} - \psi^{-\alpha}, u^\pm_i \},
\end{equation}
in which the shortened derivative,
\begin{equation}
    \bar{\Delta}^+_{\dot{\alpha}} :=
    \bar{ \partial}^+_{H\dot{\alpha}}
    =
    \frac{\partial}{\partial \bar{\theta}_H^{-\dot{\alpha}}}\,,
\end{equation}
satisfies the relations
\begin{equation}
    \delta_{lsc} \bar{\Delta} ^+_{\dot{\alpha}}
    =
    -
    \left(\bar{\Delta}^+_{\dot{\alpha}} \bar{\lambda}^{-\dot{\beta}} \right) \bar{\Delta}^+_{\dot{\beta}}\,,
    \qquad
    \bar{\Delta}^+_{\dot{\alpha}}
    \lambda^{-\beta} = 0\,.
\end{equation}
The explicit construction of the bridge \(\psi^{-\alpha}\) appearing in \eqref{Hybrid} was presented in the original paper \cite{Galperin:1987em}.
The existence of the hybrid basis  allows one to consistently formulate and resolve the covariant form of the half-analyticity constraint:
\begin{equation}\label{eq: cov half an}
    \bar{\Delta}^+_{\dot{\alpha}} \phi(Z) = 0
    \qquad
    \Rightarrow
    \qquad
    \phi(Z) =
    \phi (x_H, \theta^{+\hat{\alpha}}_H, \theta^{-\alpha}_H, u).
\end{equation}

In the hybrid basis, a few useful relations hold,
%which we present here for later use
\begin{equation}\label{eq: hebt prop}
    \begin{split}
        & (\bar{\Delta}^+)^2  \left( E  J \right) =  \left( E  J \right) \left[(\bar{\Delta}^+)^2 \left( \ln E  J \right) + 2 \bar{\Delta}_{\dot{\alpha}}   \left( \ln E  J \right)  \bar{\Delta}^{\dot{\alpha}}  \left( \ln E  J \right)  \right]  = 0,
        \\
        &(\bar{\Delta}^+)^2 \mathcal{H}^{--\alpha\dot{\alpha}, \hat{\alpha}} - 2\bar{\Delta}^{+\dot{\alpha}} \left( \ln E  J \right)  \bar{\Delta}_{\dot{\alpha}}^{+} \mathcal{H}^{--\alpha\dot{\alpha}, \hat{\alpha}} = 0,
    \end{split}
\end{equation}
where $J := \text{Ber} \frac{\partial Z_A}{\partial Z_H}$ is the Berezinian of the transformation to the hybrid basis in HSS.
These relations imply
\begin{equation}
    \begin{split}
        (\bar{\Delta}^+)^2 \left[ E J  \mathcal{H}^{--\alpha\dot{\alpha}, \hat{\alpha}}  \right]
        =
        EJ
        \Big[ & \left\{ (\bar{\Delta}^+)^2 \left( \ln E  J \right)
        +
        2 \bar{\Delta}_{\dot{\alpha}}   \left( \ln E  J \right)  \bar{\Delta}^{\dot{\alpha}}  \left( \ln E  J \right)
        \right\}
        \mathcal{H}^{--\alpha\dot{\alpha}, \hat{\alpha}}
        \\& +
        (\bar{\Delta}^+)^2 \mathcal{H}^{--\alpha\dot{\alpha}, \hat{\alpha}} - 2\bar{\Delta}^{+\dot{\alpha}} \left( \ln E  J \right)  \bar{\Delta}_{\dot{\alpha}}^{+} \mathcal{H}^{--\alpha\dot{\alpha}, \hat{\alpha}}  \Big] = 0.
    \end{split}
\end{equation}
One of the corollaries  of these relations is the vanishing of the integrals
\begin{subequations}
\begin{equation}
    \int d^4x d^8\theta du E\, X = 0,
\end{equation}
\begin{equation}\label{eq: D-- int}
    \int d^4x d^8\theta du E\,  Y^{(q)} \mathfrak{D}^{--} Z^{(2-q)} = 0\,,
\end{equation}
\end{subequations}
where $X, Y, Z$ are arbitrary covariant half-analytic superfields.

\subsection{Lorentz connection and Weyl supertensor}

 The homogeneous rigid $\mathcal{N}=2$ superconformal transformations defined in the previous Section can be represented as
 \begin{equation}\label{eq: rigid sc}
    \delta_{sc} \Phi^{(q)}_{\alpha(s_1) \dot{\alpha}(s_2)}
    =
  - s_1 \left(\mathcal{D}^+_{(\alpha} \Lambda^{-\beta)}_{sc} \right)
  \Phi^{(q)}_{\alpha(s_1-1)\beta) \dot{\alpha}(s_2)}
  -
  s_2
    \left(\bar{\mathcal{D}}^+_{(\dot{\alpha}} \bar{\Lambda}^{-\dot{\beta})}_{sc} \right)
     \Phi^{(q)}_{\alpha(s_1) \dot{\alpha}(s_2-1)\dot{\beta})}.
 \end{equation}
For superfields with only one type of indices, e.g., $\Phi^{(q)}_{\alpha(s)}$, we can consistently
impose the half-analyticity conditions:
 \begin{equation}
    \bar{\mathcal{D}}^+_{\dot{\alpha}} \Phi^{(q)}_{\alpha(s)} = 0,
    \qquad  \qquad
    \bar{\mathcal{D}}^+_{\dot{\alpha}} \left(\mathcal{D}^+_{(\alpha} \Lambda^{-\beta)}_{sc} \right)  = 0\,,
 \end{equation}
 as well as the conditions of harmonic independence,
 \begin{equation}
    \mathcal{D}^{\pm\pm} \Phi^{(q)}_{\alpha(s_1) \dot{\alpha}(s_2)}
    =
    0,
    \qquad  \qquad
    \begin{cases}
    \mathcal{D}^{\pm\pm}    \left(\mathcal{D}^+_{(\alpha} \Lambda^{-\beta)}_{sc} \right)  = 0,
    \\
        \mathcal{D}^{\pm\pm}
        \left(\bar{\mathcal{D}}^+_{(\dot{\alpha}} \bar{\Lambda}^{-\dot{\beta})}_{sc} \right)  = 0.
    \end{cases}
 \end{equation}

 Turning to local superconformal transformations, we assume that the  transformation law \eqref{eq: rigid sc} admits the natural generalization,
  \begin{equation}\label{eq: loc sc}
    \delta_{lsc} \Phi^{(q)}_{\alpha(s_1) \dot{\alpha}(s_2)}
    =
    - s_1   \Lambda_{(\alpha}^{\;\;\beta)}
    \Phi^{(q)}_{(\alpha(s_1-1)\beta) \dot{\alpha}(s_2)}
    -
    s_2
 \bar{\Lambda}_{(\dot{\alpha}}^{\;\;\dot{\beta})}
    \Phi^{(q)}_{\alpha(s_1) (\dot{\alpha}(s_2-1)\dot{\beta})},
 \end{equation}
where $ \Lambda_{(\alpha}^{\;\;\beta)}$ and its conjugate are the curved generalizations of the rotation matrices in \eqref{eq: rigid sc}.
The requirement that the half-analyticity condition \eqref{eq: cov half an} must be covariant implies
\begin{equation}
    \bar{\Delta}^+_{\dot{\alpha}} \Phi^{(q)}_{\alpha(s)} = 0,
    \qquad
    \qquad
        \bar{\Delta}^+_{\dot{\alpha}} \Lambda_{(\alpha\beta)} = 0.
\end{equation}

 Since the parameter of local conformal transformations does not satisfy the harmonic-independence condition
 (as is evident, for example, in the local Lorentz transformation sector), for the covariance of the harmonic derivatives of
 the covariant superfields \eqref{eq: loc sc} we should insert the proper \textit{ Lorentz superconnection} into the harmonic derivative:
 \begin{equation}
    \mathfrak{D}^{++}
    \quad
    \to
    \quad
    \hat{\mathfrak{D}}^{++}
    :=
    \mathfrak{D}^{++}
    +
    \mathcal{H}^{++}_{} \cdot  \mathbb{L}
    +
    \bar{\mathcal{H}}^{++}_{}
    \cdot \mathbb{R}.
 \end{equation}
 Here $\mathbb{L}$ and $\mathbb{R}$ denote generators of Lorentz group $\mathfrak{so}(1,3) \simeq \mathfrak{su}(2)_L \oplus \mathfrak{su(2)}_R$.
 The action of $\hat{\mathfrak{D}}^{++}$ on a general covariant superfield is then defined as
 \begin{equation}
        \hat{\mathfrak{D}}^{++}
        \Phi^{(q)}_{\alpha(s_1) \dot{\alpha}(s_2)}
        =
        \mathfrak{D}^{++}
        \Phi^{(q)}_{\alpha(s_1) \dot{\alpha}(s_2)}
        +
         s_1    \mathcal{H}_{(\alpha}^{++\beta)}
        \Phi^{(q)}_{(\alpha(s_1-1)\beta) \dot{\alpha}(s_2)}
        +
        s_2
        \bar{\mathcal{H}}_{(\dot{\alpha}}^{++\dot{\beta})}
        \Phi^{(q)}_{\alpha(s_1) (\dot{\alpha}(s_2-1)\dot{\beta})}\,.
 \end{equation}
 Respectively, the gauge transformations of Lorentz superconnections should be chosen as
 \begin{equation}
    \delta_{lsc} \mathcal{H}^{++}_{(\alpha\beta)}
    =
     \mathfrak{D}^{++} \Lambda_{(\alpha\beta)},
     \qquad
     \delta_{lsc} \bar{\mathcal{H}}^{++}_{(\dot{\alpha}\dot{\beta})}
     =
     \mathfrak{D}^{++} \bar{\Lambda}_{(\dot{\alpha}\dot{\beta})}.
 \end{equation}

 \medskip

Now we inspect the possible structure of the parameters $\Lambda_{(\alpha\beta)}$ and
the corresponding Lorentz superconnections $\mathcal{H}^{++}_{(\alpha\beta)}$.

Since at the linearized level we have:
\begin{subequations}
\begin{equation}
        \Lambda_{(\alpha}^{\;\;\beta)}
=
    \left(\mathcal{D}^+_{(\alpha} \Lambda^{-\beta)}_{sc} \right)
,
\qquad
    \bar{\Lambda}_{(\dot{\alpha}}^{\;\;\dot{\beta})}
=
    \left(\bar{\mathcal{D}}^+_{(\dot{\alpha}} \bar{\Lambda}^{-\dot{\beta})}_{sc} \right),
\end{equation}
\begin{equation}
       \mathcal{H}^{++}_{(\alpha\beta)} = \partial_{(\alpha}^{\dot{\beta}} G^{++}_{\beta)\dot{\beta}}
    +
    \mathcal{D}^{-}_{(\alpha} G^{+++}_{\beta)},
    \qquad
      \bar{\mathcal{H}}^{++}_{(\dot{\alpha}\dot{\beta})} = \partial_{(\dot{\alpha}}^{\beta} G^{++}_{\beta\dot{\beta})}
    +
    \bar{\mathcal{D}}^{-}_{(\dot{\alpha}} G^{+++}_{\dot{\beta})},
\end{equation}
\end{subequations}
it is natural to expect that at the full non-linear level these quantities can also be uniquely
expressed in terms of the original $\lambda$-parameters and fundamental analytic prepotentials.

The possibility of such a construction is strongly suggested
by the relation to $\mathcal{N}=2$ super-Weyl tensor. In order to define this supertensor,
we  need to include Lorentz connections also in the negatively charged harmonic derivative,
 \begin{equation}
    \mathfrak{D}^{--}
    \quad
    \to
    \quad
    \hat{\mathfrak{D}}^{--}
    :=
    \mathfrak{D}^{--}
    +
    \mathcal{H}^{--}_{} \cdot  \mathbb{L}
    +
    \bar{\mathcal{H}}^{--}_{}
    \cdot \mathbb{R},
\end{equation}
so that the zero-curvature equation \eqref{eq: SG full harm} is modified as
\begin{equation}\label{eq: SG full harm cov}
    [\hat{\mathfrak{D}}^{++} - \mathcal{H}^{(+4)}\hat{\mathfrak{D}}^{--} ,  \hat{\mathfrak{D}}^{--}]
    =
    \mathcal{D}^0.
\end{equation}
This implies the generalized harmonic ``flatness'' equations for $\mathcal{H}^{--}_{(\alpha\beta)}$:
\begin{equation}\label{eq: NL ZC equation H}
    \mathfrak{D}^{++}   \mathcal{H}^{--}_{(\alpha\beta)}
    -
    \mathfrak{D}^{--}   \mathcal{H}^{++}_{(\alpha\beta)}
    +
    \left(\mathfrak{D}^{--}   \mathcal{H}^{(+4)}\right) \mathcal{H}^{--}_{(\alpha\beta)}
    +
    2\mathcal{H}^{++\rho}_{(\alpha} \mathcal{H}^{--}_{\rho\beta)}
    =
    0,
\end{equation}
which allows one to determine $\mathcal{H}^{--}_{(\alpha\beta)}$ in terms of $\mathcal{H}^{++}_{(\alpha\beta)}$ and analytic prepotentials. The covariance
of eq. \eqref{eq: NL ZC equation H} requires the following transformation properties for $\mathcal{H}^{--}_{(\alpha\beta)}$:
\begin{equation}
    \delta_{lsc} \mathcal{H}^{--}_{(\alpha\beta)}
    =
     \mathfrak{D}^{--} \Lambda_{(\alpha\beta)},
     \qquad
     \delta_{lsc} \bar{\mathcal{H}}^{--}_{(\dot{\alpha}\dot{\beta})}
     =
     \mathfrak{D}^{--} \bar{\Lambda}_{(\dot{\alpha}\dot{\beta})}.
 \end{equation}

\medskip

Now, following \cite{Ivanov:2025gvs}, we introduce \textit{$\mathcal{N}=2$ super Weyl tensor} as
\begin{equation}\label{eq: nl Weyl}
    \mathcal{W}_{(\alpha\beta)} = (\bar{\Delta}^+)^2  \left(  EJ\, \mathcal{H}^{--}_{(\alpha\beta)} \right).
\end{equation}
It satisfies the harmonic-independence condition
\begin{equation}\label{eq: gen HI cond}
    \hat{\mathfrak{D}}^{++}   \mathcal{W}_{(\alpha\beta)}
    +
    \Gamma^{++} \mathcal{W}_{(\alpha\beta)}
 = 0,
\end{equation}
where
\begin{equation}
    \Gamma^{++}
    :=
    (-1)^M \partial_M \mathcal{H}^{++M}.
\end{equation}
This $\mathcal{N}=2$ Weyl supertensor  possesses a homogeneous local superconformal transformation law
\begin{equation}
    \delta_{lsc} \mathcal{W}_{(\alpha\beta)}
    =
    \Lambda_{(\alpha}^{\;\;\rho)}
    \mathcal{W}_{(\rho\beta)}
    +
    \Lambda_{(\beta}^{\;\;\rho)}
    \mathcal{W}_{(\alpha\rho)}
    -
    \Omega_H \mathcal{W}_{(\alpha\beta)},
\end{equation}
where
%corresponding weight factor is given by
\begin{equation}
    \Omega_H =  \partial_{\alpha\dot{\alpha}} \lambda^{\alpha\dot{\alpha}}
    -
    \partial^{-}_{\hat{\alpha}} \lambda^{+\hat{\alpha}}
    -
    \partial^{+}_{\alpha} \lambda^{-\alpha}
    +
    \partial^{--} \lambda^{++}.
\end{equation}

These reasonings imply the relation between the Lorentz connection $\mathcal{H}^{++}_{(\alpha\beta)}, \mathcal{H}^{--}_{(\alpha\beta)}$
and $\mathcal{N}=2$ Weyl supertensor $\mathcal{W}_{(\alpha\beta)}$, in terms of which the conformal supergravity constraints are solved in the covariant approach.
Indeed, solving the chirality condition for the Weyl supertensor
\begin{equation}
    \bar{\Delta}^{+}_{\dot{\alpha}} \mathcal{W}_{(\alpha\beta)} = 0
\end{equation}
as in eq. \eqref{eq: nl Weyl},
 we obtain
$\mathcal{H}^{--}_{(\alpha\beta)}$
, which is related to
$\mathcal{H}^{++}_{(\alpha\beta)}$
by the zero-curvature condition.
Though the existence of such a relation is guaranteed, its explicit form in terms of the  basic analytic prepotentials is as yet to be found.

\subsection{Higher-spin fields}

Following \cite{Ivanov:2025gvs}, we assume that in the curved case there exists a  superfield $\mathcal{H}^{++}_{\alpha(2s-2)}$
constructed in terms of spin-$\mathbf{s}$ analytic prepotential \eqref{eq: sc spin s operator} and conformal supergravity prepotentials \eqref{eq: Weyl prep},
such that it satisfies the covariant
half-analyticity condition
\begin{equation}
\Delta_{\dot{\alpha}}^+ \mathcal{H}^{++}_{\alpha(2s-2)}  = 0.
\end{equation}
 It is natural to assume that this superfield has a covariant local  transformation law \eqref{eq: loc sc}
\begin{equation}\label{eq: sg sc transf HS}
      \delta_{lsc} \mathcal{H}^{++}_{\alpha(2s-2)}
      =
       - (2s-2) \Lambda_{(\alpha}^{\;\;\beta)} \mathcal{H}^{++}_{(\beta\alpha(2s-3))}.
\end{equation}

As in the construction of the flat superspace action \eqref{eq: spin s action}, we define $\mathcal{H}^{--}_{\alpha(2s-2)}$ as a solution to
the covariant zero-curvature equation
\begin{equation}\label{eq: ZC equation}
    \hat{\mathfrak{D}}^{++} \mathcal{H}^{--}_{\alpha(2s-2)}
    -
    \hat{\mathfrak{D}}^{--} \mathcal{H}^{++}_{\alpha(2s-2)}
    +
    \left(\mathfrak{D}^{++} \mathcal{H}^{(+4)} \right)   \mathcal{H}^{--}_{\alpha(2s-2)}  = 0.
\end{equation}
The corresponding local superconformal transformations are given by
\begin{equation}
 \delta_{lsc}   \mathcal{H}^{--}_{\alpha(2s-2)}
 =
  - (2s-2) \Lambda_{(\alpha}^{\;\;\beta)} \mathcal{H}^{--}_{(\beta\alpha(2s-3))}
  -
    \left( \mathfrak{D}^{--} \lambda^{++}\right)
        \mathcal{H}^{--}_{\alpha(2s-2)}. \label{LocConf--}
\end{equation}
It is direct to check that the zero-curvature condition \eqref{eq: ZC equation} is manifestly covariant under
the local superconformal transformations \eqref{eq: sg sc transf HS} and \eqref{LocConf--}.

The appropriate generalization of the invariant action  \eqref{eq: spin s action} should then  read
\begin{equation}\label{eq: spin s lsc action}
    S^s_{\text{Weyl}} = \int d^4x d^8\theta du \, E\, \mathcal{H}^{++\alpha(2s-2)}   \mathcal{H}^{--}_{\alpha(2s-2)} .
\end{equation}
By construction, it is invariant under local superconformal transformations defined above.

The proposed action is quadratic in the spin-$\mathbf{s}$ superfields. This means  that it can be interpreted as a covariantization
of the free action of superconformal spin-$\mathbf{s}$ superfield with respect to an arbitrary supergravity background.

\subsection*{Spin-$\mathbf{s}$ gauge transformations}

Now we consider the issue of the gauge invariance of the action  \eqref{eq: spin s lsc action}.
A natural curved generalization of the flat gauge transformation law \eqref{eq: H++ gauge} reads
\begin{equation}\label{eq: gauge spin s curv}
    \begin{split}
    \delta_\lambda \mathcal{H}^{++}_{\alpha(2s-2)}
    =&\,
    \hat{\mathfrak{D}}^{++} \Lambda_{\alpha(2s-2)}
    \\
     = &\, \mathfrak{D}^{++} \Lambda_{\alpha(2s-2)}
    +
        (2s-2) \mathcal{H}^{++\beta)}_{(\alpha} \Lambda_{(\alpha(2s-3)\beta)}\,.
\end{split}
\end{equation}
It is worth to note that this form
of the gauge transformation is not obliged, in general, to coincide with the deformed
gauge transformation law for the spin-$\mathbf{s}$ supermultiplet~\eqref{eq: spin s NA mod}
derived from the hypermultiplet model \eqref{eq: 2+s}.
This means that we cannot in advance expect the invariance of the action \eqref{eq: spin s lsc action} on an arbitrary $\mathcal{N}=2$ supergravity background.

\medskip

The negatively-charged counterpart of $\mathcal{H}^{++}_{\alpha(2s-2)}$ is defined by  eq. \eqref{eq: ZC equation} and it transforms as
\begin{equation}
    \delta_\lambda \mathcal{H}^{--}_{\alpha(2s-2)} = \hat{\mathfrak{D}}^{--} \Lambda_{\alpha(2s-2)}.
\end{equation}
The gauge variation of the spin-$\mathbf{s}$ Weyl action \eqref{eq: spin s lsc action} is then given by
\begin{equation}
\delta_\lambda  S^s_{\text{Weyl}} = \int d^4x d^8\theta du \, E\,
    \left(
    \hat{\mathfrak{D}}^{++} \Lambda^{\alpha(2s-2)}
        \mathcal{H}^{--}_{\alpha(2s-2)}
    +
    \mathcal{H}^{++\alpha(2s-2)}   \hat{\mathfrak{D}}^{--} \Lambda_{\alpha(2s-2)}
     \right).
\end{equation}
The first term can be rewritten  as
\begin{equation}
    \begin{split}
    \int d^4x &d^8\theta du \, E\,
    \hat{\mathfrak{D}}^{++} \Lambda^{\alpha(2s-2)}
    \mathcal{H}^{--}_{\alpha(2s-2)}
    \\
    &=
        \int d^4x d^8\theta du \, E\,
        \left[
    \mathfrak{D}^{++} \left(  \Lambda^{\alpha(2s-2)}
    \mathcal{H}^{--}_{\alpha(2s-2)}
    \right)
    -
     \Lambda^{\alpha(2s-2)}
        \hat{\mathfrak{D}}^{++} \mathcal{H}^{--}_{\alpha(2s-2)}
    \right].
    \end{split}
\end{equation}
It can be integrated by parts, using  $\mathfrak{D}^{++} = \partial^{++} + \mathcal{H}^{++M} \partial_M$,
\begin{subequations}
\begin{equation}
    \begin{split}
        \int d^4x &d^8\theta du \, E\,
    \mathfrak{D}^{++} \left(  \Lambda^{\alpha(2s-2)}
    \mathcal{H}^{--}_{\alpha(2s-2)}
    \right)
    \\&
    =
    -   \int d^4x d^8\theta du E\,
        \left(
             \mathfrak{D}^{++} \ln E  + \Gamma^{++}
                 \right)
            \left(  \,
 \Lambda^{\alpha(2s-2)}
    \mathcal{H}^{--}_{\alpha(2s-2)}
    \right).
    \end{split}
\end{equation}
Exploiting the zero-curvature equation \eqref{eq: ZC equation}, it is easy to show that
\begin{equation}
    \begin{split}
    -   \int d^4x &d^8\theta du \, E\,
    \Lambda^{\alpha(2s-2)}
    \hat{\mathfrak{D}}^{++} \mathcal{H}^{--}_{\alpha(2s-2)}
\\& =
-   \int d^4x d^8\theta du \, E\,   \Lambda^{\alpha(2s-2)}
\left[
    \hat{\mathfrak{D}}^{--} \mathcal{H}^{++}_{\alpha(2s-2)}
    -
    \left(\mathfrak{D}^{--} \mathcal{H}^{(+4)} \right)   \mathcal{H}^{--}_{\alpha(2s-2)}
     \right].
     \end{split}
\end{equation}
\end{subequations}
Further, applying the identity
\begin{equation}
     \mathfrak{D}^{++} \ln E  + \Gamma^{++}  - \mathfrak{D}^{--} \mathcal{H}^{(+4)} = 0\,
\end{equation}
(for its derivation, see \cite{18}), we obtain
\begin{equation}
    \begin{split}
    \delta_\lambda  S^s_{\text{Weyl}}
    &=
     \int d^4x d^8\theta du \, E\,
    \left(
    -
     \Lambda^{\alpha(2s-2)}
        \hat{\mathfrak{D}}^{--}
    \mathcal{H}^{++}_{\alpha(2s-2)}
    +
    \mathcal{H}^{++\alpha(2s-2)}   \hat{\mathfrak{D}}^{--} \Lambda_{\alpha(2s-2)}
    \right)
    \\
    &=
         \int d^4x d^8\theta du \, E\,
    \left[
    -
        \mathfrak{D}^{--} \left(
    \Lambda^{\alpha(2s-2)}
    \mathcal{H}^{++}_{\alpha(2s-2)}
    \right)
    +
    2\mathcal{H}^{++\alpha(2s-2)}   \hat{\mathfrak{D}}^{--} \Lambda_{\alpha(2s-2)}
    \right].
    \end{split}
\end{equation}
In passing to the second line, we have singled out the total harmonic derivative,
which is vanishing due to \eqref{eq: D-- int}. The part of the second terms which contains no Lorentz connection
is also vanishing. However, one term still survives
\begin{equation}
    \begin{split}
        \delta_\lambda  S^s_{\text{Weyl}}
    =&
    -2 (2s-2) \int d^4x d^4\theta^+ d^2 \theta^-  \;
    \mathcal{H}^{++\alpha(2s-2)}
    (\bar{\Delta}^+)^2 \left[ EJ \mathcal{H}^{--\beta)}_{(\alpha}
     \right]   \Lambda_{\alpha(2s-3)\beta}
     \\
     =&
        -2 (2s-2) \int d^4x d^4\theta^+ d^2 \theta^-  \;
     \mathcal{H}^{++\alpha(2s-2)}
     \mathcal{W}^{\;\;\beta)}_{(\alpha}
  \Lambda_{\alpha(2s-3)\beta}.
     \end{split}
 \end{equation}

 We conclude that the invariance of the action \eqref{eq: spin s lsc action} in this form can be achieved only
 for $\mathcal{N}=2$ Weyl-flat superbackgrounds, {\it i.e.}, with  $\mathcal{W}_{(\alpha\beta)} = 0$,
 provided no additional modification of the gauge transformation law \eqref{eq: gauge spin s curv}
 and/or  the zero-curvature equations \eqref{eq: ZC equation}  has been assumed.
 This conclusion is consistent with the results obtained in the framework of the conformal superspace approach \cite{Kuzenko:2021pqm}.
 An important novelty of our formulation is that it indicates a possible way of how to modify the gauge transformations
 in order to be able to construct a gauge-invariant action on $\mathcal{N}=2$ supergravity background
 without any restrictions on the Weyl supertensor. This way implies  the successive application of the gauge transformation
 law \eqref{eq: spin s NA mod},  which finally should lead to the proper modification of the transformation law \eqref{eq: gauge spin s curv}.
 We hope to explore this possibility in more detail elsewhere.

 \medskip

 For completeness, we also present here the expression for the spin-$\mathbf{s}$ Weyl supertensor and some associate relations:
 \begin{subequations}
 \begin{equation}
    \mathcal{W}_{\alpha(2s-2)} = (\bar{\Delta}^+)^2  \left(  EJ\, \mathcal{H}^{--}_{\alpha(2s-2)} \right),
 \end{equation}
 \begin{equation}
    \hat{\mathfrak{D}}^{++}   \mathcal{W}_{\alpha(2s-2)}
    +
    \Gamma^{++}\mathcal{W}_{\alpha(2s-2)}
    = 0,
 \end{equation}
 \begin{equation}
    \delta_{lsc} \mathcal{W}_{\alpha(2s-2)}
    =
    (2s-2)\Lambda_{(\alpha}^{\;\;\beta)}
\mathcal{W}_{(\alpha(2s-3)\beta)}
    -
    \Omega_H  \mathcal{W}_{\alpha(2s-2)}.
 \end{equation}
  \end{subequations}

%\newpage

\section{Conclusions} \label{eq: sec concl}

In this paper, we have constructed the linearized off-shell superfield actions for $4D, \mathcal{N}=2$ superconformal higher-spin gauge multiplets
with an arbitrary integer spin $\mathbf{s}$ within the harmonic superspace framework.

The main features of our approach are as follows:
\begin{itemize}
    \item In order to construct the gauge-invariant $\mathcal{N}=2$ super Weyl tensors $\mathcal{W}_{\alpha(2s-2)}$ and the associate superconformal higher-spin actions,
    it proves sufficient to use only two leading supersymmetry-invariant gauge superfields, $G^{++\alpha(s-1)\dot{\alpha}(s-1)}$ and $G^{++\alpha(s-1)\dot{\alpha}(s-2)+}$,
    which specify the half-analytic superfields $\mathcal{H}^{++}_{\alpha(2s-2)}$.
    This provides an additional evidence that the extra analytic supervielbeins present from the very beginning in the superconformally covariant harmonic derivatives
    are auxiliary and serve just to ensure the possibility of  linear realization of $\mathcal{N}=2$ superconformal transformations on the hypermultiplet
    (see a similar discussion in \cite{Buchbinder:2024pjm}).

    \item Checking rigid $\mathcal{N}=2$ superconformal invariance proceeds in close analogy with the spin-$\mathbf{2}$ case \cite{Ivanov:2025gvs}
    (linearized conformal supergravity), with the superconformal symmetry being defined modulo higher-spin gauge transformations.

    \item The results obtained admit a  generalization to an arbitrary harmonic $\mathcal{N}=2$ supergravity background.
    After  such a generalization, one obtains a theory invariant under local superconformal transformations, but the gauge invariance
    is valid only on $\mathcal{N}=2$ Weyl-flat superspaces.

    \item  Ultimately, the gauge and superconformal properties of the basic superfields are determined by the standard \textit{harmonic Grassmann analyticity principle},
    whereas the structure of the action and of the $\mathcal{N}=2$ Weyl supertensors is governed by the requirement of \textit{Grassmann half-analyticity}.
    This points to the profound geometric structures that underlie
    $\mathcal{N}=2$ higher-spin superconformal theories, which are not seen in the alternative approaches.
\end{itemize}

The results presented here complete the linearized off-shell description of integer-spin $\mathcal{N}=2$ superconformal multiplets.
Among the prospective directions for the further study we distinguish:
\begin{enumerate}
    \item {\it Generalizing the harmonic superspace construction to fermionic superconformal higher-spin multiplets with half-integer spins $\mathbf{s} + \mathbf{1/2}$}.

    The simplest example of such a model for spin
    $\mathbf{3/2}$ was considered in the work \cite{Ivanov:2024bsb}. We expect that the \textit{half-analyticity principle}
    will also play the decisive role in this case;

    \item {\it Investigating nonlinear cubic and higher-order self-interactions, as well as couplings of superconformal higher-spin fields to an $\mathcal{N}=2$ supergravity background}.

    The harmonic approach suggests the following deformations of the free higher-spin gauge transformations \eqref{eq: sc spin s transformations}:
    \begin{equation}
        \delta^{(s,k)}_\lambda  \hat{\mathcal{H}}^{++}_{s} =
        \frac{1}{2} [\mathcal{D}^{++}, \hat{\mathcal{U}}_k ]
        \quad
        \to
        \quad
        \delta^{(s,k)}_\lambda  \hat{\mathcal{H}}^{++}_{s} =
        \frac{1}{2} [\mathcal{D}^{++} + \kappa_s \hat{\mathcal{H}}^{++}_{s}, \hat{\mathcal{U}}_k ].
    \end{equation}
    We hope that this modification will allow us to obtain the corresponding cubic superfield vertices in a compact geometric form.

    If the cubic interaction \eqref{eq: hyper action + sc} is generalized to an $\mathcal{N}=2$ supergravity background \eqref{eq: 2+s},
    the corresponding deformation of the flat gauge transformations is given by
    \begin{equation}
        \delta^{(s,k)}_\lambda  \hat{\mathcal{H}}^{++}_{s} =
        \frac{1}{2} [\mathcal{D}^{++}, \hat{\mathcal{U}}_k ]
        \quad
        \to
        \quad
        \delta^{(s,k)}_\lambda  \hat{\mathcal{H}}^{++}_{s} =
        \frac{1}{2} [\mathcal{D}^{++} + \kappa_2 \hat{\mathcal{H}}^{++}_{2}, \hat{\mathcal{U}}_k ],
    \end{equation}
    where the analytic operator $\hat{\mathcal{H}}^{++}_{2}$ is constructed in terms of the spin-{\bf 2} Weyl multiplet prepotentials.
    We expect that the knowledge of such a deformation (similar to the one found in ref. \cite{Beccaria:2017nco}) could help to
    determine the respective deformation of the transformation law for the composite superfield \eqref{eq: gauge spin s curv},
    as a necessary step toward placing the spin-$\mathbf{s}$ theory on generic $\mathcal{N}=2$ supergravity background.

    For further progress it would be  important to deal with the complete invariant integration measures of the full harmonic \cite{Galperin:1987ek, Galperin:1987em, Ivanov:2022vwc}
    and chiral ${\cal N}=2$ superspaces explicitly written in terms of the underlying analytic
    $\mathcal{N}=2$ supergravity prepotentials\footnote{In the approach of Refs. \cite{Kuzenko:2021pqm, Kuzenko:2024vms}
    the local superconformal transformations of coordinates
    and gauge transformations in the tangent superspace are assumed to be independent, which seems to us unnatural. The latter should be derivable from the former,
    as, e.g., in the Ogievetsky-Sokatchev formulation of conformal ${\cal N}=1$ supergravity \cite{OgSo}.}.
    It is worth noting that the half-analyticity principle generalizes to an arbitrary supergravity background \cite{Galperin:1987em} (Section \ref{sec: hybr basis}),
    and so this principle  should presumably be manifested in the structure of superconformal higher-spin theories at the full nonlinear level;

    \item {\it Analyzing induced actions for $\mathcal{N}=2$ superconformal higher-spin theories}.

    In harmonic superspace, there are well-developed methods
    to deal with the quantum effective actions \cite{Buchbinder:2026nwu}. It will be  natural to apply this powerful techniques to the higher-spin theories. In ref. \cite{Buchbinder:2024pjm}
    we accomplished the construction of the general cubic coupling of the hypermultiplet $q^{+a}$ to the whole tower of ${\cal N}=2$ integer
    higher-spin gauge multiplets and argued that the full-level gauge invariance can be achieved only with taking into account all entries of this infinite tower simultaneously.
    Treating these gauge superfields as external ones, while $q^{+a}$ as quantum, we in principle could be able to construct the
    complete effective action of the higher-spin gauge superfields just mentioned, with all higher-spin gauge invariances inherited from the cubic vertices
    $\sim \sum_{\bf s} \int d\zeta^{(-4)}q^{+a} \hat{\mathcal{H}}^{++}_{\bf s} q^+_a$.

\end{enumerate}

\acknowledgments

This work was partially supported by
the Foundation for the Advancement of Theoretical Physics and
Mathematics ``BASIS'', grant \verb|#| 25-1-1-10-4 .

\appendix

\section{Notations}\label{App: notations}

We basically follow the conventions of the book \cite{18}. The relation between spinor and vector indices is given by
\begin{equation}
    x_{\alpha\dot{\alpha}}:= \sigma^m_{\alpha\dot{\alpha}}x_m \qquad
    \partial_{\alpha\dot{\alpha}} := \frac{1}{2}\sigma^m_{\alpha\dot{\alpha}}\partial_m.
%   \qquad
%   \partial_{\alpha\dot{\beta}} \partial^{\dot{\beta}\beta}
%   = \frac{1}{4} \delta_\alpha^\beta \Box
\end{equation}

Throughout this paper, we deal with the analytic basis  $\left\{x^{\alpha\dot{\alpha}}, \theta^{\pm}_\alpha, \bar{\theta}^\pm_{\dot{\alpha}}, u_i^\pm\right\}$,
whose relation to the central basis $ \left\{x_{C}^{\alpha\dot{\alpha}}, \theta^{i}_\alpha, \bar{\theta}^i_{\dot{\alpha}}, u^\pm_i \right\}$ coordinates
is defined as:
\begin{equation}
    \begin{split}
        &x^{\alpha\dot{\alpha}} = x_C^{\alpha\dot{\alpha}} - 4i \theta^{\alpha(i} \bar{\theta}^{\dot{\alpha}j)} u^+_i u^-_j,
        \qquad
        \theta_{\alpha}^\pm = \theta^i_\alpha u^\pm_i,
        \qquad
        \bar{\theta}^\pm_{\dot\alpha}
        =
        \bar{\theta}^i_{\dot\alpha} u^\pm_i.
    \end{split}
\end{equation}
For the integration measure of the analytic superspace, we use the notation
\begin{equation}
    d \zeta^{(-4)} : = d^4x d^4\theta^+ du.
\end{equation}
We also use the chiral basis $  \left\{ x_L^{\alpha\dot{\alpha}  }, \theta^{i}_\alpha, \bar{\theta}^i_{\dot{\alpha}},  u^\pm_i  \right\},$
% when writing actions as integrals over chiral superspace
\begin{equation}
     x^{\alpha\dot{\alpha}}_L = x^{\alpha\dot{\alpha}}_C + 2i \theta^{\alpha}_i \bar{\theta}^{\dot{\alpha} i}
     =
      x^{\alpha\dot{\alpha}}
     +
     4i \theta^{-\alpha} \bar{\theta}^{+\dot{\alpha}}.
\end{equation}
The harmonic variables satisfy the condition
$u^{+i}u^-_i = 1$, which is consistent with the action of harmonic derivatives
\begin{equation}
    \partial^{++} = u^{+i} \frac{\partial}{\partial u^{-i}},
        \qquad
        \partial^{--} = u^{-i} \frac{\partial}{\partial u^{+i}},
        \qquad
        \partial^{0} = u^{+i} \frac{\partial}{\partial u^{+i}} - u^{-i} \frac{\partial}{\partial u^{-i}}.
\end{equation}
In the analytic basis, the harmonic derivatives read
\begin{equation}
    \begin{split}
    &\mathcal{D}^{++} = \partial^{++} -
    4i \theta^{+\alpha} \bar{\theta}^{+\dot{\alpha}}\partial_{\alpha\dot{\alpha}}
    +
    \theta^{+\hat{\alpha}} \partial^+_{\hat{\alpha}},
    \\
    &   \mathcal{D}^{--} = \partial^{--}
        -
    4i \theta^{-\alpha} \bar{\theta}^{-\dot{\alpha}}\partial_{\alpha\dot{\alpha}}
    +
    \theta^{-\hat{\alpha}} \partial^-_{\hat{\alpha}},
    \\
    &
        \mathcal{D}^{0} = \partial^{0}
    +
    \theta^{+\hat{\alpha}} \partial^-_{\hat{\alpha}}
    -
    \theta^{-\hat{\alpha}} \partial^+_{\hat{\alpha}}.
    \end{split}
\end{equation}
Irrespective of the basis, they satisfy the $su(2)$ relations
\begin{equation}
    [\mathcal{D}^{++}, \mathcal{D}^{--}] = \mathcal{D}^0,
    \qquad
    [\mathcal{D}^0, \mathcal{D}^{\pm\pm}] = \pm 2 \mathcal{D}^{\pm\pm}.
\end{equation}

In harmonic superspace, we deal with the superfields having a fixed harmonic $U(1)$ charge, i.e.
\begin{equation}
    \mathcal{D}^0 \Phi^{(q)} = q \Phi^{(q)} .
\end{equation}

The covariant spinor derivatives in the analytic basis are expressed as:
\begin{equation}\label{eq: cov der}
    \begin{split}
        &\mathcal{D}^+_{\hat{\alpha}} = \partial^+_{\hat{\alpha}} = (\partial^+_{{\alpha}},\;\bar{\partial}^+_{\dot{\alpha}}),
        \\
        &\mathcal{D}^-_\alpha = - \partial^-_\alpha + 4i \bar{\theta}^{-\dot{\alpha}}\partial_{\alpha\dot{\alpha}},
        \\
        &\bar{\mathcal{D}}^-_{\dot{\alpha}} = - \bar{\partial}^-_{\dot{\alpha}} - 4i \theta^{-\alpha}\partial_{\alpha\dot{\alpha}}\,,
    \end{split}
\end{equation}
and satisfy the algebra:
\begin{equation}\label{eq: rel cov derivatives}
    \begin{split} &\mathcal{D}^\pm_{\hat{\alpha}} = [\mathcal{D}^{\pm\pm}, \mathcal{D}^{\mp}_{\hat{\alpha}}],
        \\
        & [\mathcal{D}^{\pm\pm}, \mathcal{D}^\pm_{\hat{\alpha}}] = 0,
        \\& \{\mathcal{D}^+_{\alpha}, \bar{\mathcal{D}}^-_{\dot{\alpha}}\} = - \{\mathcal{D}^-_{\alpha}, \bar{\mathcal{D}}^+_{\dot{\alpha}}\}   = -4i \partial_{\alpha\dot{\alpha}}.
    \end{split}
\end{equation}

The harmonic distributions are defined by the relations \cite{18, Galperin:1985bj}
\begin{equation}\label{eq: harmonic distrib}
    \partial^{++}_1  \frac{1}{(u^+_1u^+_2)}  = \delta^{(1,-1)}(u_1, u_2),
    \qquad
    \partial^{++}_1  \frac{1}{(u^+_1u^+_2)^2}  = \partial^{--}_1 \delta^{(2,-2)}(u_1, u_2),
\end{equation}
where $\delta^{(n,-n)}(u_1, u_2)$ are the appropriate harmonic delta-functions.

\section{Half-analyticity of the superconformal gauge parameter}
\label{App: half-analytic}

In the Section 4, the superconformal variation of the half-analytic
superfield $\mathcal{H}^{++}_{\alpha(4)}$ was reduced to the law~\eqref{VARbas}
\begin{equation}\label{eq: var H Lambda}
    \delta_{sc}\mathcal{H}^{++}_{\alpha(4)}
    =
       - 4 \left(\mathcal{D}^+_{(\alpha} \Lambda_{sc}^{-\beta)} \right)
    \mathcal{H}^{++}_{(\beta\alpha(3))}
    +
    \mathcal{D}^{++}\Lambda^{(sc)}_{\alpha(4)} .
\end{equation}
Since $\mathcal{H}^{++}_{\alpha(4)}$ and $\mathcal{H}^{++}_{(\beta\alpha(3))}$ are half-analytic
and using $\bar{\mathcal{D}}^+_{\dot{\alpha}}  \left(\mathcal{D}^+_{(\alpha} \Lambda^{-\beta)}_{sc} \right) = 0$,
we obtain
\begin{equation}\label{eq: var HA}
    \delta_{sc} \left( \bar{\mathcal{D}}^+_{\dot{\alpha}} \mathcal{H}^{++}_{\alpha(4)}  \right)
    =
    \mathcal{D}^{++}
    \left(
    \bar{\mathcal{D}}^+_{\dot{\alpha}}
    \Lambda^{(sc)}_{\alpha(4)}
    \right)
    =
    0,
\end{equation}
i.e. $\mathcal{D}^{++} \Lambda^{(sc)}_{\alpha(4)} $ is a half-analytic superfield.
%Thus $\mathcal{D}^{++}\Lambda^{(sc)}_{\alpha(4)}$ is half-analytic.
Below we show that, by using the residual freedom
$\Lambda^{(sc)}\to\Lambda^{(sc)}+Z$ with $\mathcal{D}^{++}Z=0$,
one can actually make $\Lambda^{(sc)}$  half-analytic by itself.

The analysis is most transparent in the chiral basis of HSS
(see the Appendix of \cite{Ivanov:2025gvs} for details of this chiral basis).
In this basis the spinor and harmonic derivatives are short,
\begin{equation}
    \bar{\mathcal{D}}^i_{\dot{\alpha}} =
    \bar{\partial}^i_{\dot{\alpha}},
    \qquad
    \mathcal{D}^{\pm\pm}=\partial^{\pm\pm},
    \qquad
    \mathcal{D}^0=\partial^0 ,
\end{equation}
so that $\bar{\mathcal{D}}^+_{\dot{\alpha}}=u^+_i\bar{\partial}^i_{\dot{\alpha}}$
and $\mathcal{D}^{++}=\partial^{++}$ act only on $\theta$ and $u$,
respectively.

The gauge parameter has the general harmonic expansion
\begin{equation}\label{eq: harm exp}
    \Lambda^{(sc)}_{\alpha(4)}
    =
    \sum_{n=0}^\infty
    f^{i(2n)}_{\alpha(4)}
    u^+_{(i_1}\cdots u^+_{i_n}
    u^-_{i_{n+1}}\cdots u^-_{i_{2n})} ,
\end{equation}
where $f^{i(2n)}_{\alpha(4)}$ is totally symmetric and traceless.
Acting on \eqref{eq: harm exp} by $\bar{\mathcal{D}}^+_{\dot{\alpha}}$ yields
%harmonic function,
\begin{equation}\label{eq: D++ half an}
    \bar{\mathcal{D}}^+_{\dot{\alpha}}\Lambda^{(sc)}_{\alpha(4)}
    =
    \sum_{n=0}^\infty
        \left( \bar{\mathcal{D}}^i_{\dot{\alpha}} f^{i(2n)}_{\alpha(4)}
    -
    \frac{n+1}{2n+3} \bar{\mathcal{D}}_{\dot{\alpha}j} f_{\alpha(4)}^{i(2n+1)j}
    \right)
    u^+_{(i_1}\cdots u^+_{i_{n+1}}
    u^-_{i_{n+2}}\cdots u^-_{i_{2n+1})}.
\end{equation}
The explicit form of coefficients in this expansion does not matter.
What matters, is the following elementary fact of harmonic analysis:
the kernel of $\partial^{++}$ on the charge $+1$ functions is exhausted by
the level-zero term,
\begin{equation}\label{eq: kernel}
    \partial^{++}F^{(+1)}=0
    \quad\Longleftrightarrow\quad
    F^{(+1)}=A^i_{\dot{\alpha}}u^+_i ,
    \qquad
    \partial^{++}A^i_{\dot{\alpha}}=0 .
\end{equation}
Indeed, $\partial^{++}$ raises the level by one, so
$\partial^{++}F^{(+1)}=0$ forces all levels with  $n\ge1$ to vanish and
leaves us with the single $n=0$ term, the coefficient of which must be harmonic-independent.

Applying \eqref{eq: kernel} to \eqref{eq: D++ half an}, we obtain
\begin{equation}
    \bar{\mathcal{D}}^+_{\dot{\alpha}}\Lambda^{(sc)}_{\alpha(4)}
    =
       \left( \bar{\mathcal{D}}^i_{\dot{\alpha}} f_{\alpha(4)}
    -
    \frac{1}{3} \bar{\mathcal{D}}_{\dot{\alpha}j} f_{\alpha(4)}^{ij}
    \right)
    u^+_i
    :=
    A^i_{\alpha(4)\dot{\alpha}}u^+_i ,
    \qquad
    \partial^{++}A^i_{\alpha(4)\dot{\alpha}}=0 .
\end{equation}
The residual term is removed by the choice
\begin{equation}\label{eq: Z}
    Z^{(0)}_{\alpha(4)}
    =
    -\bar{\theta}^{\dot{\alpha}}_i A^i_{\alpha(4)\dot{\alpha}}
\end{equation}
so that
\begin{equation} \label{Shift}
    \bar{\mathcal{D}}^+_{\dot{\alpha}}
    \left(
    \Lambda^{(sc)}_{\alpha(4)}+Z^{(0)}_{\alpha(4)}
    \right)
    =
    0 .
\end{equation}
Moreover, since $A^i_{\alpha(4)\dot{\alpha}}$ is harmonic-independent, we have
\begin{equation}
    \partial^{++}Z^{(0)}_{\alpha(4)}
    =
    -\bar{\theta}^{\dot{\alpha}}_i
    \partial^{++}A^i_{\alpha(4)\dot{\alpha}}
    =
    0 ,
\end{equation}
so the shift \eqref{Shift} does not affect the gauge transformation:
\begin{equation}
    \mathcal{D}^{++}
    \left(
    \Lambda^{(sc)}_{\alpha(4)}+Z^{(0)}_{\alpha(4)}
    \right)
    =
    \mathcal{D}^{++}\Lambda^{(sc)}_{\alpha(4)} .
\end{equation}
Therefore one can choose the gauge parameter
$\Lambda^{(sc)}_{\alpha(4)}$ to be half-analytic from the very beginning,
\begin{equation}
    \bar{\mathcal{D}}^+_{\dot{\alpha}}\Lambda^{(sc)}_{\alpha(4)}=0 .
\end{equation}
This completes the proof of the statement after eq. \eqref{halfAnalLambda}.

%\newpage
%\appendix

%\vspace{-0.4 cm}
%\newpage


\begin{thebibliography}{99}


    \bibitem{Fradkin:1985am}
    E.S.~Fradkin and A.A.~Tseytlin,
    {\it Conformal Supergravity},
    \href{https://doi.org/10.1016/0370-1573(85)90138-3}{Phys. Rept. \textbf{119} (1985), 233-362}.
    
%\cite{Metsaev:2007rw}
\bibitem{Metsaev:2007rw}
R.~R.~Metsaev,
{\it Ordinary-derivative formulation of conformal totally symmetric arbitrary spin bosonic fields},
\href{https://doi.org/10.1007/JHEP06(2012)062}{JHEP \textbf{06} (2012), 062}
[\href{https://arxiv.org/abs/0709.4392}{arXiv:0709.4392 [hep-th]}].
%59 citations counted in INSPIRE as of 29 Oct 2026
    
    %\cite{Vasiliev:2009ck}
    \bibitem{Vasiliev:2009ck}
    M.~A.~Vasiliev,
    {\it Bosonic conformal higher-spin fields of any symmetry},
    \href{https://doi.org/10.1016/j.nuclphysb.2009.12.010}{Nucl. Phys. B \textbf{829} (2010), 176-224}
    [\href{https://arxiv.org/abs/0909.5226}{arXiv:0909.5226 [hep-th]}].
    %96 citations counted in INSPIRE as of 29 Oct 2026


    %\cite{Fradkin:1989md}
    \bibitem{Fradkin:1989md}
    E.S.~Fradkin and V.Y.~Linetsky,
    {\it Cubic Interaction in Conformal Theory of Integer Higher Spin Fields in Four-dimensional Space-time},
\href{  https://doi.org/10.1016/0370-2693(89)90120-2}{Phys. Lett. B \textbf{231} (1989), 97-106}.

    %82 citations counted in INSPIRE as of 02 Aug 2026

    %\cite{Fradkin:1989yd}
    \bibitem{Fradkin:1989yd}
    E.S.~Fradkin and V.Y.~Linetsky,
    {\it Conformal superalgebras of higher spins},
    \href{  https://doi.org/10.1016/0003-4916(90)90252-J}{Annals Phys. \textbf{198} (1990), 252-292}.
    %52 citations counted in INSPIRE as of 02 Aug 2026

    %\cite{Fradkin:1990ps}
    \bibitem{Fradkin:1990ps}
    E.S.~Fradkin and V.Y.~Linetsky,
    {\it Superconformal Higher Spin Theory in the Cubic Approximation},
\href{https://doi.org/10.1016/0550-3213(91)90262-V}{Nucl. Phys. B \textbf{350} (1991), 274-324}.

    %63 citations counted in INSPIRE as of 02 Aug 2026
    
    
    %\cite{Vasiliev:2001zy}
    \bibitem{Vasiliev:2001zy}
    M.~A.~Vasiliev,
    {\it Conformal higher spin symmetries of $4d$ massless supermultiplets and $osp(L,2M)$ invariant equations in generalized (super)space},
    \href{https://doi.org/10.1103/PhysRevD.66.066006}{Phys. Rev. D \textbf{66} (2002), 066006}
    [\href{https://arxiv.org/abs/hep-th/0106149}{arXiv:hep-th/0106149 [hep-th]}].
    %192 citations counted in INSPIRE as of 29 Oct 2026

    %\cite{Tseytlin:2002gz}
    \bibitem{Tseytlin:2002gz}
    A.A.~Tseytlin,
    {\it On limits of superstring in $AdS_5\times S^5$},
    \href{https://doi.org/10.1023/A:1020646014240}{Theor. Math. Phys. \textbf{133} (2002), 1376-1389}
    [\href{https://arxiv.org/abs/hep-th/0201112}{arXiv:hep-th/0201112 [hep-th]}].
    %156 citations counted in INSPIRE as of 17 Aug 2026


    \bibitem{Segal:2002gd}
    A.Y.~Segal,
    {\it Conformal higher spin theory},
    \href{https://doi.org/10.1016/S0550-3213(03)00368-7}{Nucl. Phys. B \textbf{664} (2003), 59-100},
    [\href{https://arxiv.org/abs/hep-th/0207212}{arXiv:hep-th/0207212 [hep-th]}].

    %\cite{Bekaert:2010ky}
    \bibitem{Bekaert:2010ky}
    X.~Bekaert, E.~Joung, and J.~Mourad,
    {\it Effective action in a higher-spin background},
    \href{https://doi.org/10.1007/JHEP02(2011)048}{JHEP \textbf{02} (2011), 048}
    [\href{https://arxiv.org/abs/1012.2103}{arXiv:1012.2103 [hep-th]}].
    %141 citations counted in INSPIRE as of 17 Aug 2026

    %\cite{Bekaert:2022poo}
    \bibitem{Bekaert:2022poo}
    X.~Bekaert, N.~Boulanger, A.~Campoleoni, M.~Chiodaroli, D.~Francia, M.~Grigoriev, E.~Sezgin, and E.~Skvortsov,
    {\it Snowmass White Paper: Higher Spin Gravity and Higher Spin Symmetry},
    [\href{https://arxiv.org/abs/2205.01567}{arXiv:2205.01567 [hep-th]}].
    %96 citations counted in INSPIRE as of 17 Aug 2026


    %\cite{Beccaria:2017nco}
    \bibitem{Beccaria:2017nco}
    M.~Beccaria and A.~A.~Tseytlin,
    {\it On induced action for conformal higher spins in curved background},
    \href{ https://doi.org/10.1016/j.nuclphysb.2017.03.022}{Nucl. Phys. B \textbf{919} (2017), 359-383}
    [\href{https://arxiv.org/abs/1702.00222}{arXiv:1702.00222 [hep-th]}].
    %33 citations counted in INSPIRE as of 30 Sep 2026

    %\cite{Basile:2022nou}
    \bibitem{Basile:2022nou}
    T.~Basile, M.~Grigoriev, and E.~Skvortsov,
    {\it Covariant action for conformal higher spin gravity},
    \href{https://doi.org/10.1088/1751-8121/aceeca}{J. Phys. A \textbf{56} (2023) no.38, 385402}
    [\href{https://arxiv.org/abs/2212.10336}{arXiv:2212.10336 [hep-th]}].
    %20 citations counted in INSPIRE as of 17 Aug 2026




    %\cite{Kuzenko:2017ujh}
    \bibitem{Kuzenko:2017ujh}
    S.M.~Kuzenko, R.~Manvelyan, and S.~Theisen,
    {\it Off-shell superconformal higher spin multiplets in four dimensions},
\href{https://doi.org/10.1007/JHEP07(2017)034}{JHEP \textbf{07} (2017), 034}
    [\href{https://arxiv.org/abs/1701.00682}{arXiv:1701.00682 [hep-th]}].
    %65 citations counted in INSPIRE as of 17 Aug 2026

    %\cite{Kuzenko:2019ill}
    \bibitem{Kuzenko:2019ill}
    S.M.~Kuzenko and M.~Ponds,
    {\it Conformal geometry and (super)conformal higher-spin gauge theories},
    \href{https://doi.org/10.1007/JHEP05(2019)113}{JHEP \textbf{05} (2019), 113}
    [\href{https://arxiv.org/abs/1902.08010}{arXiv:1902.08010 [hep-th]}].
    %41 citations counted in INSPIRE as of 17 Aug 2026

    %\cite{Kuzenko:2020jie}
    \bibitem{Kuzenko:2020jie}
    S.M.~Kuzenko, M.~Ponds, and E.S.N.~Raptakis,
    {\it New locally (super)conformal gauge models in Bach-flat backgrounds},
    \href{https://doi.org/10.1007/JHEP08(2020)068}{JHEP \textbf{08} (2020), 068}
    [\href{https://arxiv.org/abs/2005.08657}{arXiv:2005.08657 [hep-th]}].
    %27 citations counted in INSPIRE as of 17 Aug 2026

    %\cite{Kuzenko:2020opc}
    \bibitem{Kuzenko:2020opc}
    S.M.~Kuzenko, M.~Ponds, and E.S.N.~Raptakis,
    {\it Generalised superconformal higher-spin multiplets},
    \href{https://doi.org/10.1007/JHEP03(2021)183}{JHEP \textbf{03} (2021), 183}
    [\href{https://arxiv.org/abs/2011.11300}{arXiv:2011.11300 [hep-th]}].
    %11 citations counted in INSPIRE as of 17 Aug 2026

    %\cite{Kuzenko:2022hdv}
    \bibitem{Kuzenko:2022hdv}
    S.~M.~Kuzenko, M.~Ponds, and E.S.N.~Raptakis,
    {\it Conformal Interactions Between Matter and Higher-Spin (Super)Fields},
    \href{https://doi.org/10.1002/prop.202200157}{Fortsch. Phys. \textbf{71} (2023) no.1, 1}
    [\href{https://arxiv.org/abs/2208.07783}{arXiv:2208.07783 [hep-th]}].
    %20 citations counted in INSPIRE as of 17 Aug 2026

    %\cite{Kuzenko:2022qeq}
    \bibitem{Kuzenko:2022qeq}
    S.~M.~Kuzenko, J.~La Fontaine, and M.~Ponds,
    {\it Induced action for superconformal higher-spin multiplets using SCFT techniques},
    \href{  https://doi.org/10.1016/j.physletb.2023.137818}{Phys. Lett. B \textbf{839} (2023), 137818}
    [\href{https://arxiv.org/abs/2212.00468}{arXiv:2212.00468 [hep-th]}].
    %10 citations counted in INSPIRE as of 17 Aug 2026


    %\cite{Kuzenko:2021pqm}
    \bibitem{Kuzenko:2021pqm}
    S.M.~Kuzenko and E.S.N.~Raptakis,
    {\it Extended superconformal higher-spin gauge theories in four dimensions},
    \href{https://doi.org/10.1007/JHEP12(2021)210}{JHEP \textbf{12} (2021), 210}
    [\href{https://arxiv.org/abs/2104.10416}{arXiv:2104.10416 [hep-th]}].
    %34 citations counted in INSPIRE as of 02 Aug 2026

    %\cite{Kuzenko:2024vms}
    \bibitem{Kuzenko:2024vms}
    S.M.~Kuzenko and E.S.N.~Raptakis,
    {\it Towards $ \mathcal{N}=2$ superconformal higher-spin theory},
    \href{https://doi.org/10.1007/JHEP11(2024)013}{JHEP \textbf{11} (2024), 013}
    [\href{https://arxiv.org/abs/2407.21573}{arXiv:2407.21573 [hep-th]}].
    %1 citations counted in INSPIRE as of 31 Dec 2024


    %\cite{Hutchings:2023iza}
    \bibitem{Hutchings:2023iza}
    D.~Hutchings, S.M.~Kuzenko, and E.S.N.~Raptakis,
    {\it The $\mathcal{N}=2$ superconformal gravitino multiplet},
    \href{https://doi.org/10.1016/j.physletb.2023.138132}{Phys. Lett. B \textbf{845} (2023), 138132}
    [\href{https://arxiv.org/abs/2305.16029}{arXiv:2305.16029 [hep-th]}].
    %6 citations counted in INSPIRE as of 14 Nov 2024

%\cite{Galperin:1984av}
\bibitem{Galperin:1984av}
A.~Galperin, E.~Ivanov, S.~Kalitzin, V.~Ogievetsky, and E.~Sokatchev,
{\it Unconstrained $\mathcal{N}=2$ Matter, Yang-Mills and Supergravity Theories in Harmonic Superspace},
\href{https://doi.org/10.1088/0264-9381/1/5/004}{Class. Quant. Grav. \textbf{1} (1984), 469-498}
[erratum: \href{https://doi.org/10.1088/0264-9381/2/1/512}{Class. Quant. Grav. \textbf{2} (1985), 127}].
%848 citations counted in INSPIRE as of 14 Nov 2024

\bibitem{18} A.S.~Galperin, E.A.~Ivanov, V.I.~Ogievetsky, E.S.~Sokatchev,
{\it Harmonic superspace}, Cambridge Monographs on Mathematical
Physics, Cambridge University Press, 2001, 306 p.

\bibitem{GRAn}
A.S.~Galperin, E.A.~Ivanov, and V.I.~Ogievetsky, {\it Grassmann analyticity and extended supersymmetries}, JETP Lett. \textbf{33} (1981), 176-181.
   %\cite{Galperin:1985zv}

%2 citations counted in INSPIRE as of 08 Jan 2023









%\cite{Buchbinder:2021ite}
\bibitem{Buchbinder:2021ite}
I.~Buchbinder, E.~Ivanov, and N.~Zaigraev,
{\it Unconstrained off-shell superfield formulation of $4D,  \mathcal{N}=2$ supersymmetric higher spins},
\href{https://doi.org/10.1007/JHEP12(2021)016}{JHEP \textbf{12} (2021), 016}
[\href{https://arxiv.org/abs/2109.07639}{arXiv:2109.07639 [hep-th]}].
%19 citations counted in INSPIRE as of 14 Nov 2024

%\cite{Buchbinder:2022kzl}
\bibitem{Buchbinder:2022kzl}
I.~Buchbinder, E.~Ivanov, and N.~Zaigraev,
{\it Off-shell cubic hypermultiplet couplings to $\mathcal{N}=2$ higher spin gauge superfields},
\href{https://doi.org/10.1007/JHEP05(2022)104}{JHEP \textbf{05} (2022), 104}
[\href{https://arxiv.org/abs/2202.08196}{arXiv:2202.08196 [hep-th]}].
%17 citations counted in INSPIRE as of 14 Nov 2024


%\cite{Buchbinder:2022vra}
\bibitem{Buchbinder:2022vra}
I.~Buchbinder, E.~Ivanov, and N.~Zaigraev,
{\it $\mathcal{N}=2$ higher spins: superfield equations of motion, the hypermultiplet supercurrents, and the component structure},
\href{https://doi.org/10.1007/JHEP03(2023)036}{JHEP \textbf{03} (2023), 036}
[\href{https://arxiv.org/abs/2212.14114}{arXiv:2212.14114 [hep-th]}].
%9 citations counted in INSPIRE as of 14 Nov 2024



%\cite{Buchbinder:2024pjm}
\bibitem{Buchbinder:2024pjm}
I.~Buchbinder, E.~Ivanov, and N.~Zaigraev,
{\it $\mathcal{N}=2$ superconformal higher-spin multiplets and their hypermultiplet couplings},
\href{https://doi.org/10.1007/JHEP08(2024)120}{JHEP \textbf{08} (2024), 120}
[\href{https://arxiv.org/abs/2404.19016}{arXiv:2404.19016 [hep-th]}].
%4 citations counted in INSPIRE as of 14 Nov 2024

%\cite{Buchbinder:2024xll}
\bibitem{Buchbinder:2024xll}
I.~Buchbinder, E.~Ivanov, and N.~Zaigraev,
{\it $\mathcal{N} = 2$ higher spin theories and harmonic superspace},
\href{https://doi.org/10.22323/1.455.0048}{PoS \textbf{ICPPCRubakov2023} (2024), 048}
[\href{https://arxiv.org/abs/2402.05704}{arXiv:2402.05704 [hep-th]}].
%5 citations counted in INSPIRE as of 17 Aug 2026

%\cite{Buchbinder:2025ceg}
\bibitem{Buchbinder:2025ceg}
I.~Buchbinder, E.~Ivanov, and N.~Zaigraev,
{\it Towards $\mathcal{N}=2$ higher-spin supergravity},
[\href{https://arxiv.org/abs/2503.02438}{arXiv:2503.02438 [hep-th]}].
%1 citations counted in INSPIRE as of 02 May 2025

%\cite{Ivanov:2024gjo}
\bibitem{Ivanov:2024gjo}
E.~Ivanov and N.~Zaigraev,
{\it Off-shell invariants of linearized $4D,\mathcal{N}=2$ supergravity in the harmonic approach},
\href{https://doi.org/10.1103/PhysRevD.110.066020}{Phys. Rev. D \textbf{110} (2024) no.6, 066020}
[\href{https://arxiv.org/abs/2407.08524}{arXiv:2407.08524 [hep-th]}].
%1 citations counted in INSPIRE as of 14 Nov 2024

%\cite{Ivanov:2025mld}
\bibitem{Ivanov:2025mld}
E.A.~Ivanov and N.M.~Zaigraev,
{\it $\mathcal{N} = 2$ Supergravity and Harmonic Superspace: Linearized Supercurvatures},
\href{https://doi.org/10.1134/S106377962470148X}{Phys. Part. Nucl. \textbf{56} (2025) no.2, 247-252}.



\bibitem{Ivanov:2024bsb}
E.~Ivanov and N.~Zaigraev,
{\it $\mathcal{N}=2$ superconformal gravitino in harmonic superspace},
\href{https://doi.org/10.1016/j.physletb.2025.139333}{Phys. Lett. B \textbf{862} (2025), 139333}
[arXiv:2412.14822 [hep-th]].

%\cite{Ivanov:2025gvs}
\bibitem{Ivanov:2025gvs}
E.~Ivanov and N.~Zaigraev,
{\it Linearized $\mathcal{N}=2$ conformal supergravity in the harmonic approach},
\href{https://doi.org/10.1007/JHEP09(2026)270}{JHEP \textbf{09} (2026), 270}
[\href{https://arxiv.org/abs/2511.16325}{arXiv:2511.16325 [hep-th]}].
%2 citations counted in INSPIRE as of 10 Jul 2026


%\cite{Zaigraev:2024ryg}
\bibitem{Zaigraev:2024ryg}
N.~Zaigraev,
{\it $\mathcal{N}=2$ higher-spin supercurrents},
\href{https://doi.org/10.1016/j.physletb.2024.139056}{Phys. Lett. B \textbf{858} (2024), 139056}
[\href{https://arxiv.org/abs/2408.00668}{arXiv:2408.00668 [hep-th]}].
%8 citations counted in INSPIRE as of 06 Aug 2026


%\cite{Zaigraev:2025xqo}
\bibitem{Zaigraev:2025xqo}
N.M.~Zaigraev,
{\it $\mathcal{N} = 2$ Abelian Interactions and Higher-Spin Supercurrents},
\href{https://doi.org/10.1134/S1547477125701754}{Phys. Part. Nucl. Lett. \textbf{22} (2025) no.6, 1374-1377}.
%1 citations counted in INSPIRE as of 17 Aug 2026

%\cite{Zaigraev:2026jvu}
\bibitem{Zaigraev:2026jvu}
N.~Zaigraev,
{\it Structure of $\mathcal{N} = 2$ superfield higher-spin abelian cubic interactions},
\href{https://doi.org/10.1007/JHEP09(2026)181}{JHEP \textbf{09} (2026), 181}
[\href{https://arxiv.org/abs/2605.27206}{arXiv:2605.27206 [hep-th]}].
%2 citations counted in INSPIRE as of 06 Aug 2026


%\cite{Zaigraev:2026xzr}
\bibitem{Zaigraev:2026xzr}
N.~Zaigraev,
{\it Novel $ \mathcal{N} $ = 2 higher-spin supercurrents},
\href{https://doi.org/10.1007/JHEP09(2026)074}{JHEP \textbf{09} (2026), 074}
[\href{https://arxiv.org/abs/2606.05370}{arXiv:2606.05370 [hep-th]}].
%1 citations counted in INSPIRE as of 06 Aug 2026

%\cite{Zaigraev:2026vxf}
\bibitem{Zaigraev:2026vxf}
N.~Zaigraev,
{\it On Born-Infeld-type higher-spin interactions and supersymmetry},
[\href{https://arxiv.org/abs/2610.31652}{arXiv:2610.31652 [hep-th]}].
%0 citations counted in INSPIRE as of 27 Oct 2026


%\cite{Galperin:1987ek}
\bibitem{Galperin:1987ek}
A.S.~Galperin, E.A.~Ivanov, V.I.~Ogievetsky, and E.~Sokatchev,
{\it $\mathcal{N}=2$ Supergravity in Superspace: Different Versions and Matter Couplings},
\href{https://doi.org/10.1088/0264-9381/4/5/023}{Class. Quant. Grav. \textbf{4} (1987), 1255}.
%96 citations counted in INSPIRE as of 14 Nov 2024

\bibitem{Galperin:1985zv}
A.~Galperin, E.~Ivanov, V.~Ogievetsky, and E.~Sokatchev,
{\it Conformal invariance in harmonic superspace},
[\href{https://lib-extopc.kek.jp/preprints/PDF/1985/8508/8508325.pdf}{Dubna preprint JINR E2-85-363}], (1985) , published in "Quantum Field
theory and Quantum Statistics", eds. I. Batalin, C. J. Isham, G. Vilkovisky, vol.2,
Adam Hilger, Bristol, 1987, p. 233–248.



%\cite{Ivanov:2005qf}
\bibitem{Ivanov:2005qf}
E.A.~Ivanov, A.V.~Smilga, and B.M.~Zupnik,
{\it Renormalizable supersymmetric gauge theory in six dimensions},
\href{https://doi.org/10.1016/j.nuclphysb.2005.08.014}{Nucl. Phys. B \textbf{726} (2005), 131-148}
[\href{https://arxiv.org/abs/hep-th/0505082}{arXiv:hep-th/0505082 [hep-th]}].
%85 citations counted in INSPIRE as of 16 Sep 2026

%\cite{Buchbinder:2020yvf}
\bibitem{Buchbinder:2020yvf}
I.L.~Buchbinder, E.A.~Ivanov, and V.A.~Ivanovskiy,
{\it Superfield realization of hidden $R$-symmetry in extended supersymmetric gauge theories and its applications},
\href{https://doi.org/10.1007/JHEP04(2020)126}{JHEP \textbf{04} (2020), 126}
[\href{https://arxiv.org/abs/2001.01649}{arXiv:2001.01649 [hep-th]}].
%3 citations counted in INSPIRE as of 16 Sep 2026

%\cite{Zaigraev:2026gjb}
\bibitem{Zaigraev:2026gjb}
N.~Zaigraev and J.~Zernina,
{\it Wess-Zumino gauge for analytic prepotentials of off-shell $\mathcal{N}=2$ supergravity: bosonic sector},
[\href{https://arxiv.org/abs/2608.27308}{arXiv:2608.27308 [hep-th]}].
%0 citations counted in INSPIRE as of 29 Sep 2026

%\cite{Ivanov:2024bsb}
%1 citations counted in INSPIRE as of 02 May 2025

%\cite{Galperin:1987em}
\bibitem{Galperin:1987em}
A.S.~Galperin, N.A.~Ky, and E.~Sokatchev,
{\it $\mathcal{N}=2$ Supergravity in Superspace: Solution to the Constraints},
\href{https://doi.org/10.1088/0264-9381/4/5/022}{Class. Quant. Grav. \textbf{4} (1987), 1235}.
%55 citations counted in INSPIRE as of 14 Nov 2024

%\cite{Ivanov:2022vwc}
\bibitem{Ivanov:2022vwc}
E.~Ivanov,
{\it $\mathcal{N}=2$ Supergravities in Harmonic Superspace}, in: Bambi, C., Modesto, L., Shapiro, I. (eds) \href{https://doi.org/10.1007/978-981-19-3079-9\_43-1}{Handbook of Quantum Gravity, Springer, 2024}
[\href{https://arxiv.org/abs/2212.07925}{arXiv:2212.07925 [hep-th]}].
%7 citations counted in INSPIRE as of 04 May 2025




%\cite{Zupnik:1998td}
%\bibitem{Zupnik:1998td}
%B.~M.~Zupnik,
%{\it Background harmonic superfields in $\mathcal{N}=2$ supergravity},
%\href{https://doi.org/10.1007/BF02557138}{Theor. Math. Phys. \textbf{116} (1998), 964-977}
%[\href{https://arxiv.org/abs/hep-th/9803202}{arXiv:hep-th/9803202 [hep-th]}].
%10 citations counted in INSPIRE as of 21 Nov 2024



%\cite{Galperin:1985bj}
\bibitem{Galperin:1985bj}
A.~Galperin, E.A.~Ivanov, V.~Ogievetsky, and E.~Sokatchev,
{\it Harmonic Supergraphs. Green Functions},
\href{https://doi.org/10.1088/0264-9381/2/5/004}{Class. Quant. Grav. \textbf{2} (1985), 601}.
%211 citations counted in INSPIRE as of 19 Mar 2026


\bibitem{OgSo}
V.~Ogievetsky and E.~Sokatchev,
{\it Structure of Supergravity Group},
\href{https://doi.org/10.1016/0370-2693(78)90228-9}{Phys. Lett. B \textbf{79} (1978), 222}.
%148 citations counted in INSPIRE as of 16 Sep 2026


%\cite{Buchbinder:2026nwu}
\bibitem{Buchbinder:2026nwu}
I.L.~Buchbinder and E.A.~Ivanov,
{\it Supersymmetry at BLTP: Recent Progress in Two Directions},
\href{https://doi.org/10.54546/NaturalSciRev.200707}{Natural Sci. Rev. \textbf{3} (2026), 200707}
[\href{https://arxiv.org/abs/2606.25063}{arXiv:2606.25063 [hep-th]}].
%2 citations counted in INSPIRE as of 22 Sep 2026

\end{thebibliography}
\end{document}